\documentclass[final,1p,times]{elsarticle}

\usepackage{amssymb}
\usepackage{amsmath}
\usepackage{amsthm}

\journal{Annals of Physics}

\begin{document}

\begin{frontmatter}



\title{Inflationary Universe in Deformed Phase Space Scenario}


\author[1,2]{S. M. M. Rasouli}
\author[4]{Nasim Saba}
\author[4]{Mehrdad Farhoudi}
\author[1,2]{Jo\~ao Marto}
\author[1,2]{P.V. Moniz}

\address[1]{Departamento de F\'{i}sica,
Universidade da Beira Interior, Rua Marqu\^{e}s d'Avila
e Bolama, 6200-001 Covilh\~{a}, Portugal.}
\address[2]{Centro de Matem\'{a}tica e Aplica\c{c}\~{o}es (CMA - UBI),
Universidade da Beira Interior, Rua Marqu\^{e}s d'Avila
e Bolama, 6200-001 Covilh\~{a}, Portugal.}

\address[4]{Department of Physics, Shahid Beheshti University, G.C.
             Evin, Tehran, 19839, Iran.}

\begin{abstract}
We consider a noncommutative (NC) inflationary model with a
homogeneous scalar field minimally coupled to gravity.
The particular NC inflationary setting herein proposed, produces
entirely new consequences as summarized in what follows.
We first analyze the free field case and subsequently examine
the situation where the scalar field is subjected to a polynomial and exponential potentials.
We propose to use a canonical deformation between momenta, in a spatially
flat Friedmann-Lema\^{\i}tre-Robertson-Walker~(FLRW) universe,
and while the Friedmann equation (Hamiltonian constraint) remains unaffected
the Friedmann acceleration equation (and thus the Klein-Gordon equation) is
modified by an extra term linear in the NC parameter. This concrete noncommutativity
 on the momenta allows interesting dynamics that other NC models seem not to allow. Let us be more precise.
This extra term behaves as the sole explicit pressure that under the right circumstances
implies a period of accelerated expansion of the universe.
We find that in the absence of the scalar field potential, and in contrast with the
commutative case, in which the scale factor always
decelerates, we obtain an inflationary phase for small negative
values of the NC parameter. Subsequently, the period of accelerated expansion
is smoothly replaced by an appropriate deceleration phase providing
an interesting model regarding the graceful exit problem in inflationary models.
This last property is present either in the free field case or under the influence of
the scalar field potentials considered here. Moreover, in the case of the
free scalar field, we show that not only the horizon problem is solved but also there
 is some resemblance between the evolution equation of the scale factor associated to
 our model and that for the $R^2$ (Starobinsky) inflationary model.
Therefore, our herein NC model not only can be taken as an appropriate scenario to get a successful kinetic inflation,
but also is a convenient setting to obtain inflationary universe possessing the graceful exit when scalar field
potentials are present.

\end{abstract}


\end{frontmatter}


\section{Introduction}
\label{int}
\indent

As Einstein gravitational theory is not suitable to describe the
universe at very high energies,
alternative proposals must provide an essential new perspective.
In this regard, Snyder's
formulation~\citep{SNY,SNY1} for a NC setting of spacetime
coordinates is of significant interest.
It  introduces a short length cutoff (that is
called the NC parameter) which can modify the renormalizability
properties of relativistic quantum field theory (see~\citep{noncom1,noncom} and references therein
for a thorough review;
 cf., e.g.,~\citep{IRUV,TRN,TRN1,HO1,LV,LV2,HO4,LV3,phrev3,HAS1,GW,GW1,MF12,
RFK11,phrev1,RZMM14,RM14,MF21,RZJM16,sf2016} for several specific explorations.) At the scales where quantum gravity effects would be
important, NC effects could therefore be
relevant\footnote{String/M theories have
added interest into the framework discussion regarding noncommutativity,  due to the natural appearance
of  NC spacetime~\citep{STMT,STMT2,STMT1} (see
also~\citep{DFR,DFR0,DFR1,DFR2} and references therein). More precisely,
  the spacetime uncertainty relation $\Delta t\Delta x>l^{2}_{s}$
(where $t$ and $x$ are physical time and space coordinates, respectively, and $l_{s}$
is the string length scale), emerging in string/M theories, indicates that
the spacetime could be noncommutative at particular scales~\citep{inf00}.}.
In particular, as  inflation proceeds from  such energy scales, employing
deformed phase space scenarios for investigating this dynamical stage of the universe is surely pertinent.  Accordingly, we may expect that such
correction from the spacetime uncertainty principle (implying a deviation from
general relativity) may affect the cosmic microwave background
power spectrum and hence may be identified in future cosmological observational data.

As far as the inflationary paradigm framework currently stands\footnote{As a brief summary for the benefit of a reader, let us just add that inflationary models have been considered to overcome unsatisfactory
aspects of the (standard) hot big bang scenario,  such as the flatness, horizon
and monopole problems~\citep{gu,li,al,li1,linde}.
It has been believed that the horizon problem is a quite acute problem with respect to
 the flatness and the monopole issues,  such that every
number of e-foldings that can resolve the horizon problem,
automatically, can solve the others as well~\citep{wein2}.
Moreover,
such models also allow to predict the primordial power spectrum of
the density fluctuations, which are presumed to be a seed for all
the observed structure in the universe~\citep{6}. However, in any  satisfactory inflationary model, the universe must exit from the
rapidly accelerating phase and then proceed into a decelerating expansion stage (where
normal baryonic matter, radiation, neutrinos and dark matter
become dominant). Such a transition is often called the graceful exit and still constitutes a
well-known problem to be faced by any inflationary model.},
it has been widely acquiesced that a scalar field (usually designated as  the inflaton),  is the responsible for
the period of accelerated expansion during that earliest epoch of the
universe.

In the original proposal (of inflation) by Guth \citep{gu}, it was assumed that a scalar field is trapped in a
false vacuum. Subsequently, by tunneling through a
quantum-mechanical barrier, it is possible for the inflaton field to exit from this
local minimum value. Then, via a first order transition, it can go towards a
true vacuum associated to the present universe.
However, in this hypothetical process,  inflation cannot terminate successfully.
In order to overcome this problem, a new inflationary model had been independently proposed by
Linde~\citep{li}, Alberch and Steinhard~\citep{al}, which is indeed a
modified version of the aforementioned scenario. In the new inflationary model, the
inflaton field varies slowly in a double-well potential and
undergoes a phase transition of the second order.
In this slow evolutionary behavior associated to the scalar field, its corresponding potential
energy dominates its own kinetic energy; via such an assumption  the universe expands
quasi-exponentially, associated to  slow-roll approximation (SRA) conditions.
In this setting, the inflationary epoch terminates when the potential
energy stops dominating.
The simplest example of the new
inflationary model is the chaotic inflation in which the potential, with a sufficiently flat
region, sustain a slow-roll
regime~\citep{li1}.

Let us mention that noncommutativity has been employed regarding inflation in
the recent literature. Concretely, in Ref.~\citep{inf4}, it has been shown that a NC spacetime affected power law
inflation and could provide a large enough running of the spectra index; in Ref.~\citep{inf0},
the effect of noncommutativity on cosmic microwave background has been
investigated and it has been shown that noncommutativity may cause the
spectrum of fluctuation to be non-Gaussian and anisotropic.
Moreover, in Ref.~\citep{inf1,inf2,inf3,inf4}, the effects of NC spacetime
on the power spectrum, spectral index and running spectral index of the
curvature perturbations have been investigated in the inflationary universe.
However, notwithstanding as well the content in
Refs.~\citep{inf0,inf00,inf1,inf2,inf3,inf4,inf5}, where several
types of NC frameworks have been proposed
to study the early universe, we use instead a rather different
NC relation in our herein model, as we will explain in our manuscript.

In this paper, we investigate the effects of a particular type of
noncommutativity with a spatially flat
Friedmann-Lema\^{\i}tre-Robertson-Walker~({FLRW})
cosmological model
in the presence of a scalar field which is minimally
coupled to gravity.
It has been proposed that the geometry of the universe at early times
was not commutative, namely, we should incorporate imprints of the NC geometry
in the description of the universe at those times \citep{AMMOO05}.
Therefore, our purpose is to study the effects
of that specific NC property (associated
to the conjugate momentum sector) regarding inflationary scenarios.
Employing our NC model, in which just one NC parameter is showing up linearly
in the equations of motion, we show that, even in the case of free scalar field
(in the absence of the scalar potential),
for very small values of the NC parameter, there is a short epoch at early times in
which the universe inflates. Subsequently, the universe enters in a
decelerating era which can be considered as the radiation dominated epoch.
We should note such a phase transition behavior can never be obtained in the commutative
case, where the scale factor of the universe always decelerates.
Moreover, in order to overcome to the main problem with the standard cosmology, i.e., the horizon problem,
we show that in our NC kinetic inflationary universe, the relevant
nominal condition is completely satisfied during the evolution of the universe.
Furthermore, we discuss briefly regarding the close similarity between the herein NC
inflation and the $R^2$ (Starobinsky) inflationary model \citep{S80,V85} at the level of the equation
associated to the evolution of the scale factor and demonstrate
that we can find more relevance for interpreting the NC models in very small scales.
However, for the free scalar field case, the commutative model does not
yield an accelerating phase, nor does it satisfy the nominal condition.
In the presence of the scalar potential, we extend the standard SRA setting and
employ this procedure for a few well known scalar potentials.
Using small values for the NC parameter, we show that the
noncommutativity affects in the values of the numbers
of e-folding as well as in behavior of the slow-roll parameters, scalar
field and the Hubble parameter depicted versus the logarithmic scale factor.

This work is organized as follows.
In the next section, by employing the Hamiltonian
formalism and proposing a particular kind of a dynamical
deformation between the conjugate momentum sector,
we obtain the corresponding NC field equations.
In section \ref{free potential}, we analyze our model in the absence of
the scalar potential and compare the results with those obtained from the commutative case.
In section~\ref{sec-SRA}, by employing the SRA procedure, we obtain
a generalized set of the SRA parameters/relations for our herein NC model
for a class of polynomial potentials.
Subsequently, in section~\ref{numerical}, with the assistance  of numerical analysis, using
re-scaled variables and choosing a  set of suitable initial conditions,
we present  the effects of our chosen NC deformation in the field equations and cosmological observables.
Finally, in the last section, we summarize the main results and present a short discussion.

\section{Noncommutative Cosmological Scenario}
\label{Standard}

Our background spacetime is
described by the spatially flat FLRW universe
\begin{equation}\label{frw}
ds^{2}=-\mathcal{N}^2(t)dt^2+a^2(t)\left(dx^2+dy^2+dz^2\right),
\end{equation}
where $t$ is the cosmic time and $x, y, z$ are the Cartesian
coordinates; $\mathcal{N}(t)$ is a lapse function and $a(t)$ is the scale
factor.
We employ the well known Lagrangian density
\begin{equation}
{\mathcal{L}}=\sqrt{-g}\left[\frac{1}{16\pi G}R-\frac{1}{2}g^{\mu\nu}\phi_{,\mu}\phi_{,\nu}-V(\phi)\right],
\label{eq1}
\end{equation}
in which the scalar field $\phi$ is minimally
coupled to gravity. In (\ref{eq1}), $g$ is the determinant of the metric
$g_{\mu\nu}$, $R$ is the Ricci scalar, $V(\phi)$ is a scalar potential,
 the Greek indices run from zero to three and we have assumed the
 units in which $c=1$ and $\hbar=1$.

By inserting the expression for the Ricci scalar [associated to the metric~(\ref{frw})] into Lagrangian~(\ref{eq1}), we get
\begin{equation}\label{Lag}
{\mathcal{L}}=-\frac{3}{8\pi G}\mathcal{N}^{-1}a\dot{a}^2+\frac{1}{2}\mathcal{N}^{-1}a^3\dot{\phi}^2-\mathcal{N}a^3V,
\end{equation}
where an overdot represents a derivative with respect to $t$ and we have omitted the total time derivative term.
It can be shown that the Hamiltonian of the model is given by~\citep{RZMM14}
\begin{equation}\label{H0}
\mathcal{H}=-\frac{2}{3}\pi G \mathcal{N}a^{-1}P_{a}^{2}+\frac{1}{2}\mathcal{N}a^{-3}P_{\phi}^{2}+\mathcal{N}a^{3}V(\phi),
\end{equation}
where $P_{a}$ and $P_{\phi}$ are the momenta conjugates associated
to the scale factor and the scalar field, respectively.
We henceforth take the comoving gauge where $\mathcal{N}=1$.
The equations of motion associated to the commutative case (correspond to the phase
space coordinates $\{a,\phi;P_a,P_\phi\}$) are obtained by considering the ordinary
phase space structure in which
\begin{equation}\label{notdeformed}
\{a,P_{a}\}=1\qquad {\rm and}\qquad   \{\phi,P_{\phi}\}=1
\end{equation}
and the other brackets vanish.
Consequently, employing the corresponding Hamiltonian~(\ref{H0}), we get
\begin{eqnarray}
\dot{a}\!\!\! & = & \!\!\!\{a,{\mathcal{H}}\}=-\frac{4\pi G}{3}a^{-1}P_{a},\label{diff.eq1}\\
\dot{P}_{a}\!\!\! & = & \!\!\!\{P_{a},{\mathcal{H}}\}=
-\frac{2\pi G}{3}a^{-2}P_{a}^{2}+\frac{3}{2}a^{-4}P_{\phi}^{2}-3a^{2}V(\phi),\label{diff.eq2}\\
\dot{\phi}\!\!\! & = & \!\!\!\{\phi,{\mathcal{H}}\}=a^{-3}P_{\phi},\label{diff.eq3}\\
\dot{P}_{\phi}\!\!\! & = &
\!\!\!\{P_{\phi},{\mathcal{H}}\}=-a^{3}V'(\phi),\label{diff.eq4}
\end{eqnarray}
where the prime represents the derivative with respect to the argument.

As far as our NC setting is regarded, let us elaborate on it from hereafter. The literature includes, concerning canonical deformation by means of Poisson brackets,   either the
Moyal product (i.e.,  the star-product, see, e.g., \citep{nonocomshrefs1} and references therein) or the generalized
uncertainty principle~\citep{GUPrefs}.  Applying these and other NC frameworks (at  classical or quantum levels)
into cosmological settings, have enabled to explore
important challenges.
For instance, it allowed to reasonably address UV/IR mixing, as means to describe and relate in a non-trivial
manner
physical phenomena at large and short distances (or equivalently, high and low energy regimes). This has been achieved as an  outcome of employing  NC
quantum field theories; see, e.g., Refs.~\citep{reviewnoncom1,minwramsei}.
Consequently, importing   NC features into  (classical or quantum) cosmology can be soundly motivated as opening quite promising avenues to explore.
In the present work, we shall restrict
ourselves to a classical geometrical framework, where the corresponding NC effects will be
 obtained by using classical canonical noncommutativity features into Poisson brackets.

Let us therefore  employ a specific type of a canonical  noncommutativity, which is
obtained by means of an appropriate deformation on the classical phase space variables.
Our choice has been seldom \citep{RZMM14} used in the literature, namely in inflationary settings but has computational advantages.
We will explain that the corresponding equations of
motion (associated to the deformed scenario) can still be obtained
by employing the Hamiltonian (\ref{H0}), being
evaluated on variables which satisfy the  deformed Poisson
bracket. Therefore, let us apply the deformed Poisson bracket between
the canonical conjugate momenta as\footnote{In this work (see also footnote 7), we have used the units where
$\hbar=1=c$, therefore, from the Planck length, $L_{\rm P}=\sqrt{\hbar
G/c^3}$, the dimension of $G$ is $[G]=L_{\rm P}^2$. We assumed that the scale factor and the lapse function to be
dimensionless parameters, and the dimensions of coordinates and the
Lagrangian are $[x^\mu]=L_{\rm P}$ and $[{\cal L}]=L_{\rm P}^{-4}$. Consequently, we can
show that $[\phi]=L_{\rm P}^{-1}$, $[P_a]=L_{\rm P}^{-3}$, $[P_\phi]=L_{\rm P}^{-2}$, and
consequently $[\{P_a,P_\phi\}]=L_{\rm P}^{-2}$. Therefore, assuming~(\ref{deformed}) yields the
dimension of the deformation parameter as $[\theta]=L_{\rm P}$.}
\begin{equation}\label{deformed}
\{P_{a},P_{\phi}\}=\theta\phi^{3},
\end{equation}
where the NC parameter $\theta$ has been assumed as a constant.
We should note that it is also possible to assume other choices
for the right hand side of~(\ref{deformed}), still
satisfying the dimensionality of $\{P_a,P_\phi\}$, but the present suggestion~(\ref{deformed})
reveals to be particularly interesting because it is linear (most simple dependence) in terms of the
deformation parameter and the NC parameter does not appear
in the Friedmann equation (Hamiltonian constraint).
In addition, more motivations concerning noncommutativity between the momenta can be found in~\citep{RFK11}.
Moreover, employing the NC ingredient~(\ref{deformed})
for studying the gravitational collapse of a homogeneous scalar field produced interesting results~\citep{RZMM14}. It is worthwhile to note that if in instead of the NC Poisson bracket (\ref{deformed}) for momenta, we used a NC upon only the scale factor and the scalar field, then any NC effects will be absent  for a
vanishing potential \citep{GSS11}.
However, an important outcome in our model, is that with (\ref{deformed}) we still get
modified field equations  for the case where the scalar potential is absent as we will elaborate about.
In summary, we believe that this dynamical noncommutativity between the momenta provides
more interesting dynamics to describe the evolution of the universe, at least in the early times, than
other choice of modified Poisson brackets.

Before proceeding, let us just clarify  that
the  phase space structure (\ref{notdeformed}) is still employed , with the
modified configuration being brought  from the relation (\ref{deformed}), which is the sole
responsible as a canonical NC feature; it  induces a set of modified equations as the novel  framework to explore, as we will elaborate in the following.

It is then straightforward to show that the modified
equations of motion with respect to the
Hamiltonian~(\ref{H0}) are given by
\begin{eqnarray}
\dot{P}_{a}\!\!\! & = & -\frac{2\pi G}{3}a^{-2}P_{a}^{2}+\frac{3}{2}a^{-4}P_{\phi}^{2}-3a^{2}V(\phi)+\theta \left(a^{-3}\phi^{3}P_{\phi}\right),\label{diff.eq-prime2}\\
\dot{P}_{\phi}\!\!\! & = &\!\!-a^{3}V'(\phi)+\theta\left(\frac{4\pi G}{3}
a^{-1}\phi^{3}P_{a}\right).\label{diff.eq-prime4}
\end{eqnarray}
We note that as equations (\ref{diff.eq1}) and
(\ref{diff.eq3}), under the chosen noncommutativity, are not
modified, we have forborne from rewriting them.
Moreover, in order to obtain equations (\ref{diff.eq-prime2}) and (\ref{diff.eq-prime4}), we have
employed the following formulas\footnote{In~\citep{RZMM14}, two different
approaches have been used to retrieve the equations of motion.}
\begin{eqnarray}\label{calc}
\{P_a,f(P_a,P_\phi)\}=\theta\phi^3\frac{\partial f}{\partial P_\phi},\\
\{P_\phi,f(P_a,P_\phi)\}=-\theta\phi^3\frac{\partial f}{\partial P_a}.
\end{eqnarray}
Obviously, the standard commutative equations are recovered in the limit
$\theta\rightarrow 0$.

The equations of motion
associated to our herein NC framework can be written as the standard form as
\begin{eqnarray}\label{asli1}
H^{2}=\frac{8\pi G}{3}\left(\frac{1}{2}
\dot{\phi}^{2}+V(\phi)\right)\equiv\frac{8\pi G}{3}\rho_{\rm tot},
\end{eqnarray}
\begin{eqnarray}\label{asli2}
2\frac{\ddot{a}}{a}+H^{2}=-8\pi G\left[\left(\frac{1}{2}\dot{\phi}^{2}-V(\phi)\right)+\frac{\theta\phi^{3}\dot{\phi}}{3a^2}\right]\equiv-8\pi Gp_{\rm tot},
\end{eqnarray}
\begin{equation}\label{asli3}
\ddot{\phi}+3H\dot{\phi}+V'(\phi)+\theta H\left(\frac{\phi^3}{a^2}\right)=0,
\end{equation}
where $H\equiv\dot{a}/a$ is the Hubble parameter and we have
employed the Hamiltonian constraint $\mathcal{H}=0$.

Moreover, in this NC model, the energy density
and pressure associated to the scalar field have been
denoted by $\rho_{\rm tot}$ and $p_{\rm tot}$, respectively.
Let us also introduce $p_{\rm nc}\equiv\frac{\theta\phi^{3}\dot{\phi}}{3a^2}$, which denotes the
sole {\it explicit} term representing the direct NC  effects in the total pressure.
We should note that not only the $p_{\rm nc}$ {\it explicitly} depends
on the NC parameter, but also the two first terms of $p_{\rm tot}$ as well as the $\rho_{\rm tot}$
{\it implicitly} depend on the NC parameter.
We emphasize that there is no appropriate manner to separate the commutative portion unless setting $\theta=0$.

Therefore, in analogy with standard cosmology, the equation of state can be written as
\begin{equation}\label{w}
w_{\rm tot}=\frac{p_{\rm tot}}{\rho_{\rm tot}}=
\frac{\dot{\phi}^{2}-2V(\phi)+\frac{2\theta\phi^{3}\dot{\phi}}{3a^2}}{\dot{\phi}^{2}+2V(\phi)}.
\end{equation}
If we set $\theta=0$ (here and in the field equations),
we get the same equation of state associated to the
standard models, namely,  $w_{\rm tot}=w_{\rm \phi}$,
where $w_{\rm \phi}<-1/3$, which corresponds to $\dot{\phi}^{2}<V(\phi)$, being associated with the
quintessence cosmological model for the late times. Whilst, a dominant potential
energy with respect to the kinetic term, can lead to an inflationary epoch at very early times.

By using the conservation equation
\begin{equation}\label{cons}
\dot{\rho}_{\rm tot}+3H(\rho_{\rm tot}+p_{\rm tot})=0,
\end{equation}
 time derivative of the
Hubble parameter (that will be needed later on) is
\begin{equation}\label{dh}
\dot{H}=-4\pi G\left(\dot{\phi}^2+\frac{\theta\phi^{3}\dot{\phi}}{3a^{2}}\right).
\end{equation}
As expected, let us repeat again, in all of the above equations if we set $\theta=
0$, then, each equation reduces to its corresponding commutative
counterpart.

In addition, one evident but very pertinent impact of the NC deformation
studied in this work is that the NC dependent terms in Eqs.~(\ref{asli2})-(\ref{dh})
are, at least, proportional to the inverse square of the scale factor $a(t)$. Therefore,
it is expected that the NC effect should be noticeable at the initial stage of inflation
and very residual at its end.

\section{Kinetic inflation and the horizon problem}
\label{free potential}

In this section, we want to present and analyze the NC effects when
the scalar field potential is absent and compare them with those found from the standard framework.

By assuming $V=0$, from (\ref{asli1}), it is easy to show that the scale factor is related to the scalar field as
\begin{equation}\label{exp-rel}
a(t)=a_0e^{\kappa\phi(t)},
\end{equation}
where $a_0>0$ is an integration constant and $\kappa=\pm\sqrt{\frac{4\pi G}{3}}=\pm\frac{\sqrt{2}}{2}l$,
where $l=4.7\times10^{-33}cm$ is the Planck length and we should take the
positive (upper) sign. Moreover, in this section, we work with the units where $8\pi G=1$. Furthermore, relation (\ref{exp-rel}) yields
\begin{equation}\label{exp-rel-2}
H=\kappa\dot{\phi}(t).
\end{equation}



\begin{figure}
\includegraphics[width=2.5in]{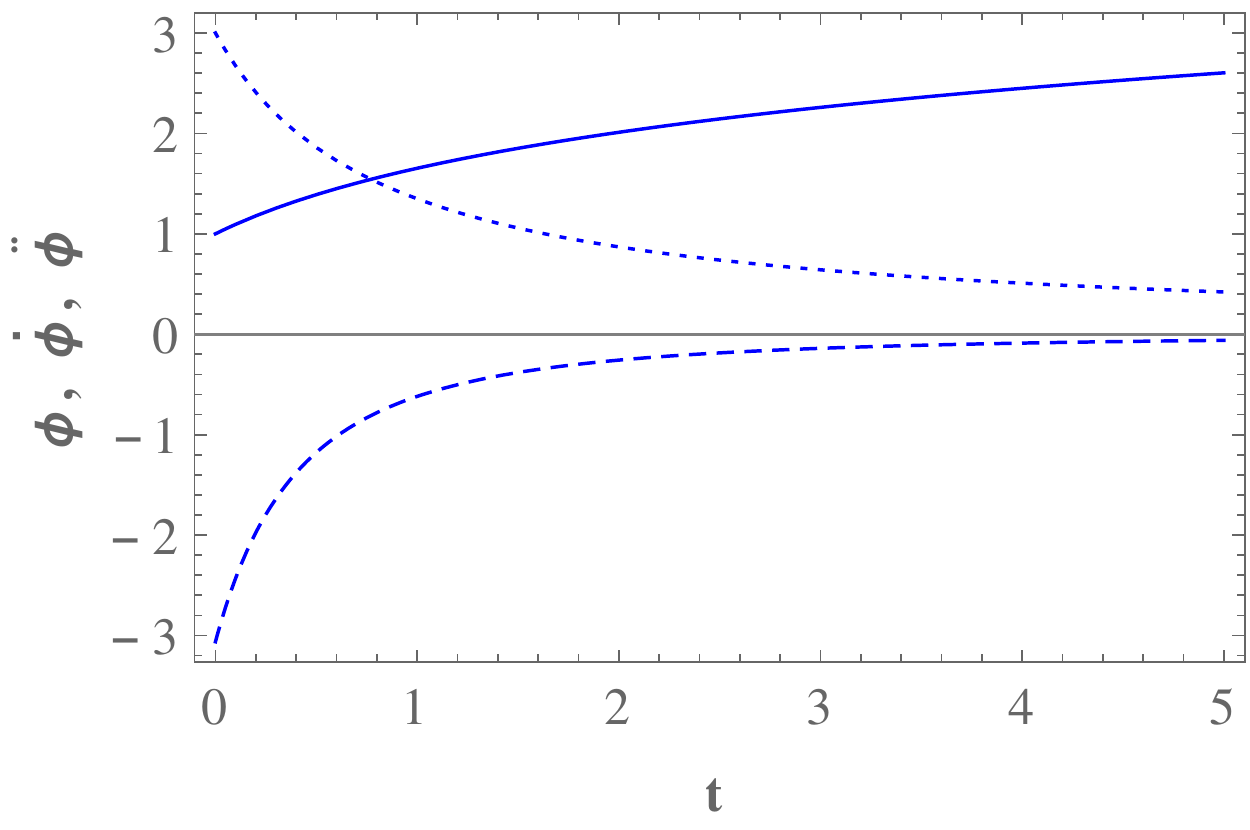}
\includegraphics[width=2.5in]{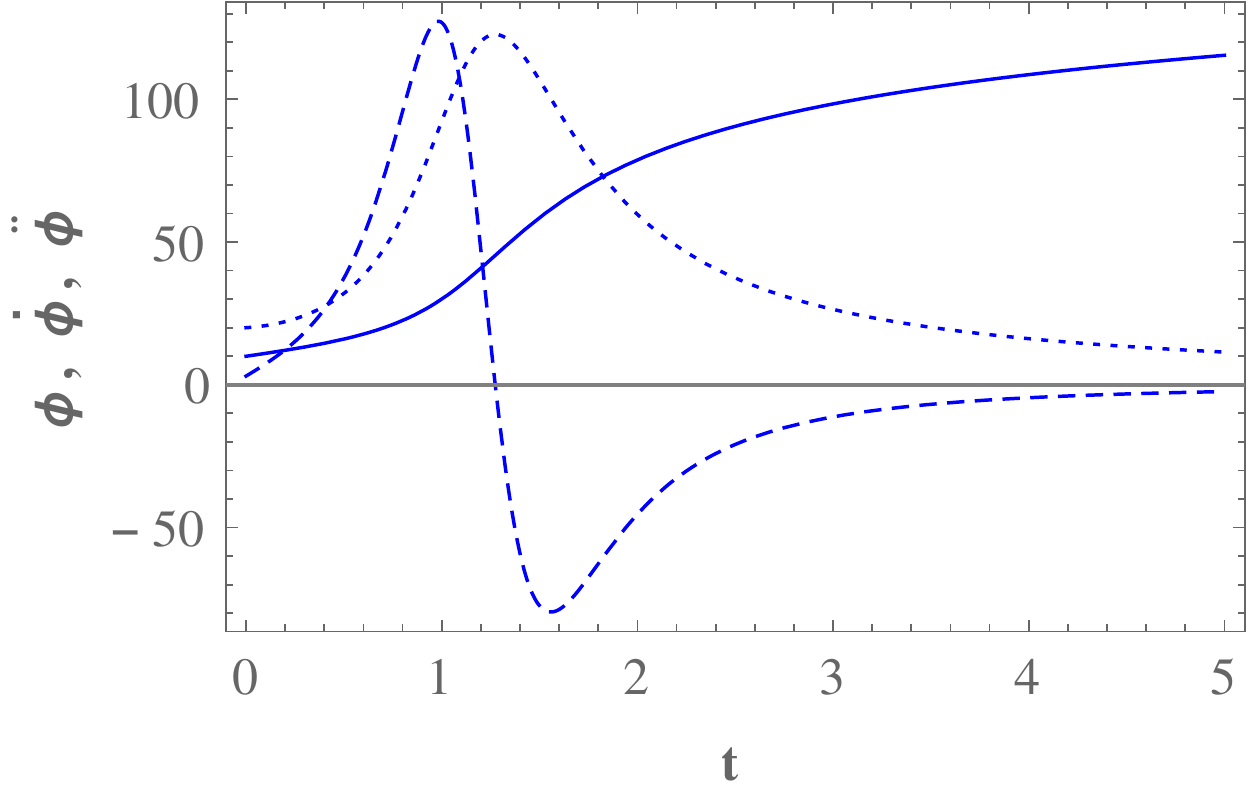}
\caption{The behavior of $\phi(t)$ (solid curves), $\dot{\phi}(t)$ (dotted curves) and $\ddot{\phi}(t)$ (dashed curves)
 against cosmic time for commutative case (left panel) and for NC case (right panel). Moreover, we suggest to reduce the
 The red line is associated to $\phi(t)=0$ to clearly show when those quantities are positive or negative.
We have set $8\pi G=1$, $\kappa=\frac{\sqrt{6}}{6}$, $a_0=0.01$ ,
$\phi(0)=1=\dot{\phi}(0)$ and $\theta=-0.0008$ for the NC case.
 For more clarity, we have re-scaled the curves.}
\label{phi-zero}
\end{figure}

\begin{figure}
\includegraphics[width=2.5in]{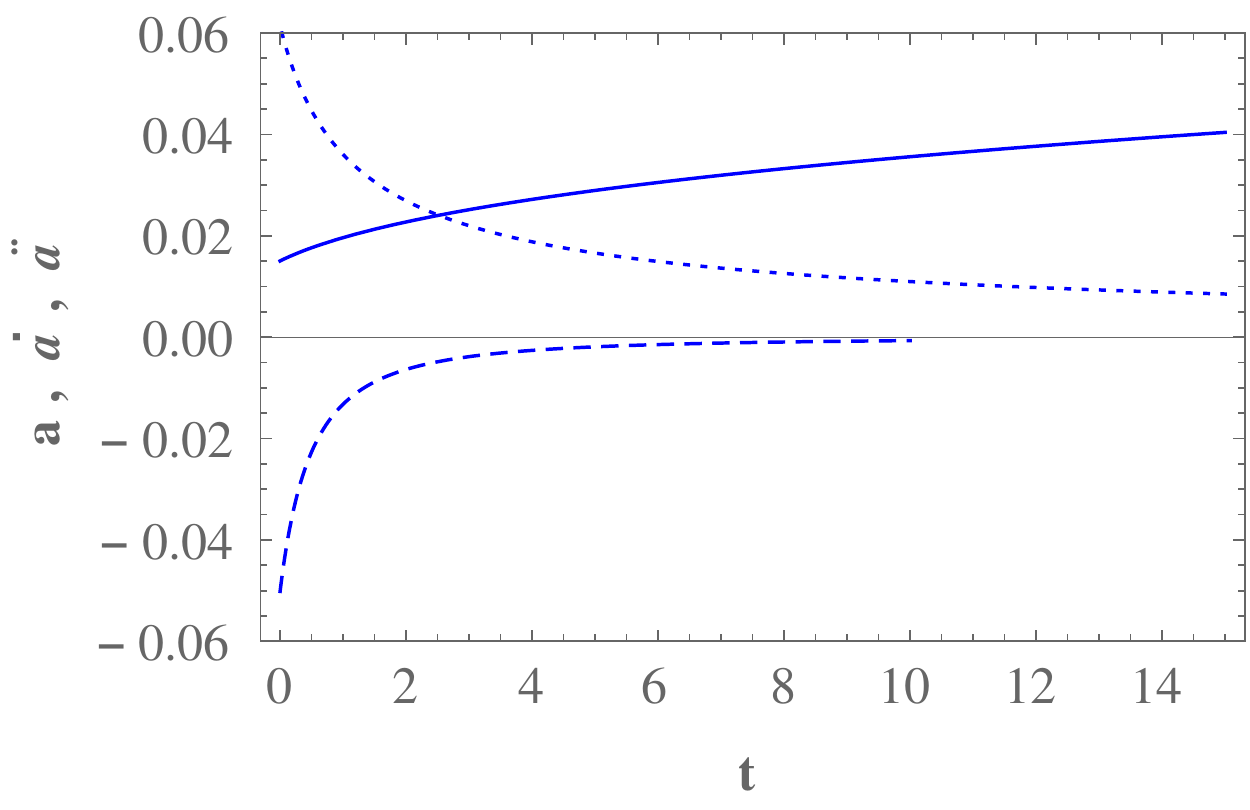}
\includegraphics[width=2.5in]{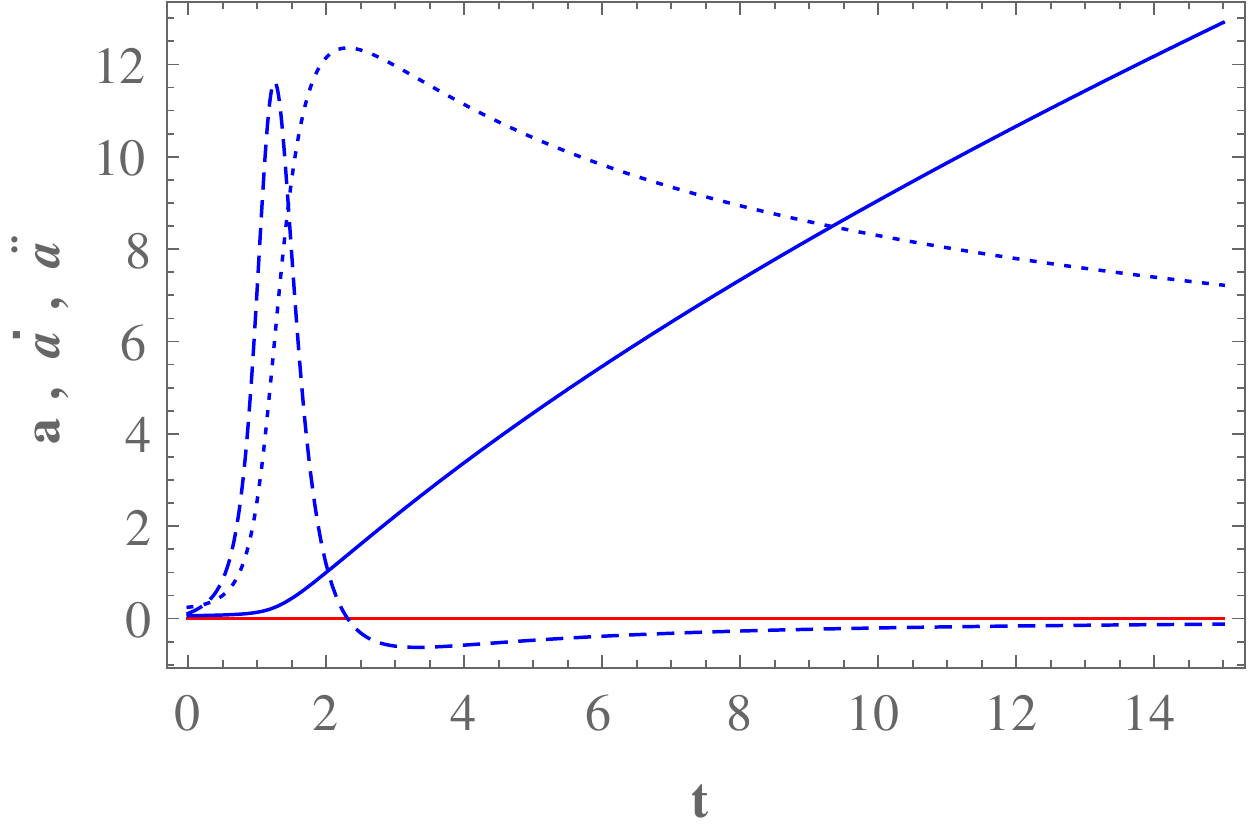}
\caption{The behavior of $a(t)$ (solid curves), $\dot{a}(t)$ (dotted curves) and $\ddot{a}(t)$ (dashed curves)
 against cosmic time for commutative case (left panel) and for NC case (right panel).
 The red line in
 the lower panel is associated to $a(t)=0$ to clearly show when those quantities are positive or negative.
We have set $8\pi G=1$, $\kappa=\frac{\sqrt{6}}{6}$, $a_0=0.01$ ,
$\phi(0)=1=\dot{\phi}(0)$ and $\theta=-0.0008$ for the NC case.
 For more clarity, we have re-scaled  the curves.}
\label{a-zero}
\end{figure}

\begin{figure}
\centering\includegraphics[width=3.2in]{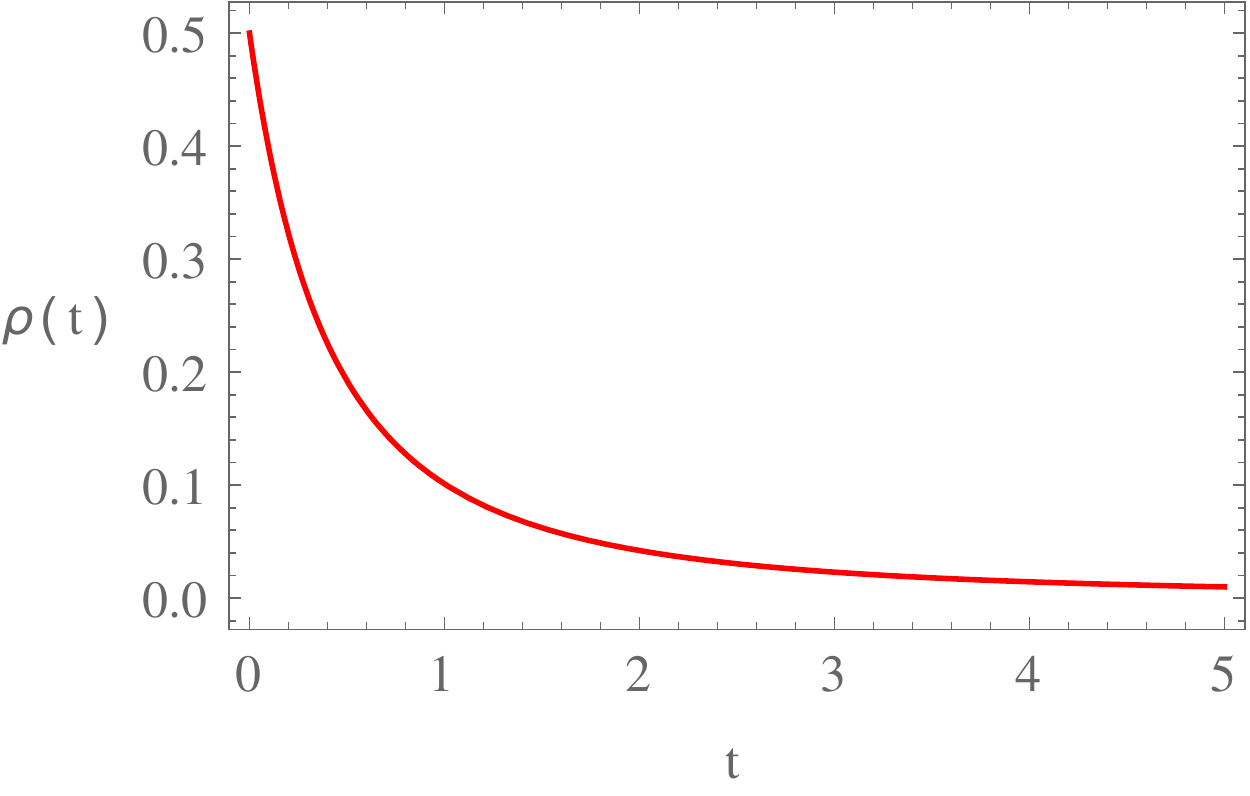}\\
\centering\includegraphics[width=3.2in]{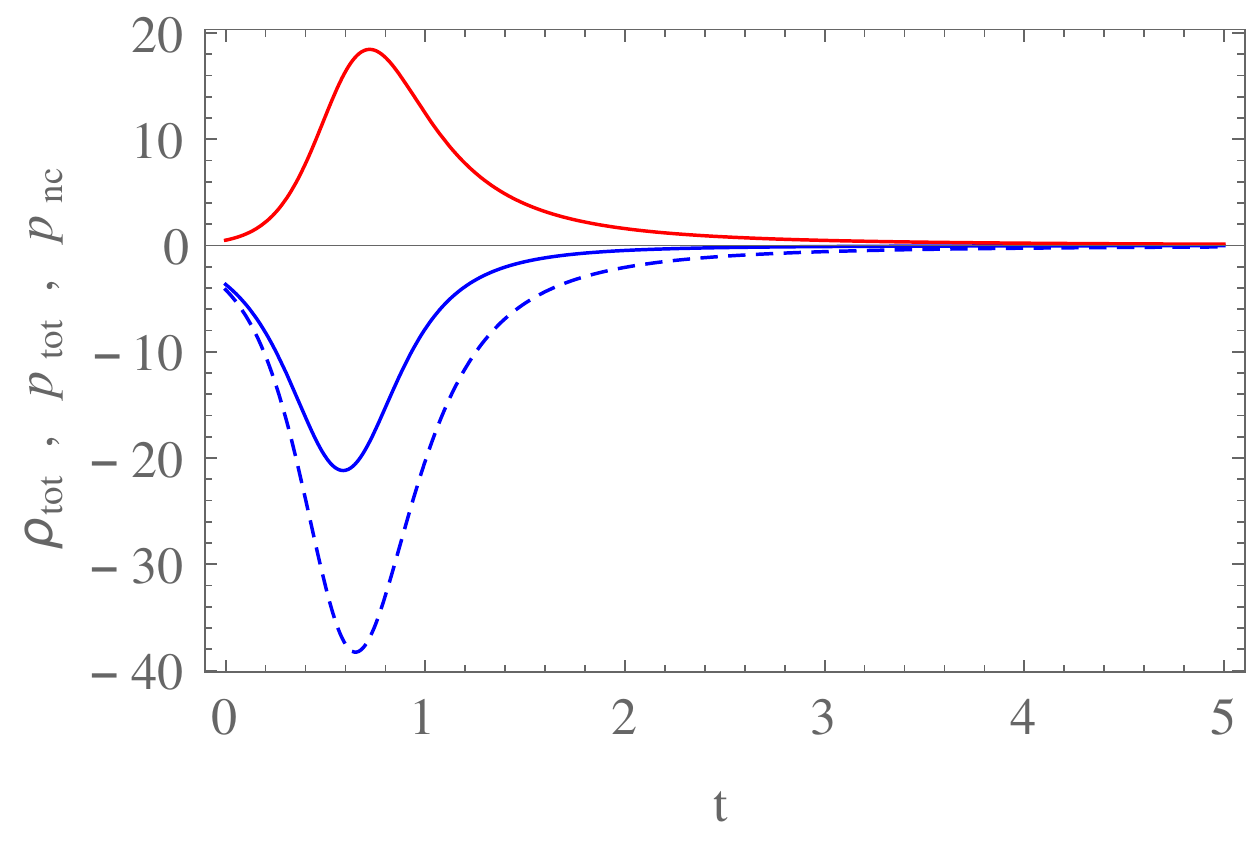}
\centering\includegraphics[width=3.2in]{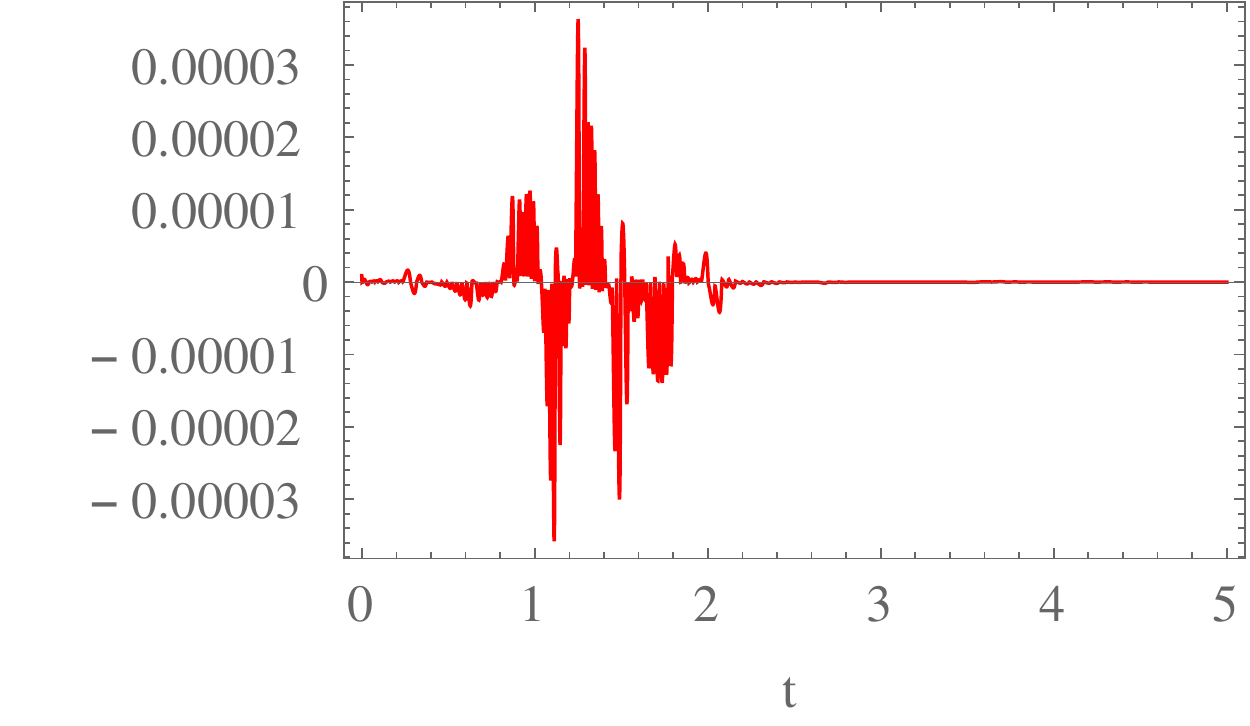}
\caption{The behavior of the kinetic energy (red curves), total pressure (solid blue curve)
and NC pressure (dashed blue curve) against cosmic time for
commutative case (upper panel) and for NC case (middle panel).
The lower panel shows the numerical error for the time evolution associated to satisfying the conservation equation (\ref{cons}).
We have set $8\pi G=1$, $\kappa=\frac{\sqrt{6}}{6}$, $a_0=0.01$ , $\phi(0)=1=\dot{\phi}(0)$ and $\theta=-0.0008$ for the NC case.}
\label{rho-p-tot-zero}
\end{figure}

 By substituting (\ref{exp-rel}) and (\ref{exp-rel-2}) into the
 modified Klein-Gordon equation (\ref{asli3}), we write
 \begin{equation}\label{c-zero-V-1}
\ddot{\phi}+3\kappa\dot{\phi}^2+\kappa\theta \left(\frac{\dot{\phi}\phi^3}{a_0^2e^{2\kappa\phi(t)}}\right)=0.
\end{equation}
In order to discuss the NC consequences within our model, contrasting with those obtained
from a standard cosmological scenario (in the absence of potential), let us obtain
the solution associated to the commutative case.
It is straightforward to show that, for $\theta=0$, we get
the following relations for the scalar field and scale factor \citep{SS17}
\begin{eqnarray}\label{c-zero-V-2}
\phi(t)=\frac{1}{3\kappa}{\rm ln}\left[3\kappa(c_1t+t_0)\right],\hspace{10mm}
a(t)=a_0\left[3\kappa(c_1t+t_0)\right]^{\frac{1}{3}},
\end{eqnarray}
 where $c_1>0$ and $t_0>0$ are integration constants and $t>-\frac{t_0}{c_1}$.
 Let us in what follows describe the behavior of the quantities
 associated to this case. At $t\rightarrow-t_0/c_1$, we get $\phi\rightarrow-\infty$; the scalar field
 increases with the cosmic time and goes to $+\infty$ for very large times.
 Moreover, for all times we see that $\frac{\ddot{\phi}}{\phi}<0$, namely, the scalar field always decelerates.
 Therefore, the energy decreases with the cosmic time and tends to zero
 after an infinite expansion. In addition,  the scale factor of the universe starts
 its decelerating expansion from a nonzero value.
 Concerning the time behavior of quantities associated to the
 commutative case, we present a few examples, by using particular initial conditions, in
  figures \ref{phi-zero}, \ref{a-zero} and \ref{rho-p-tot-zero} (the upper panels).

For $\theta\neq0$, solving the complicated differential equation (\ref{c-zero-V-1})
analytically is impossible.
However, it is feasible to derive the
general conditions under which the universe can accelerate.
Moreover, we can obtain a condition concerning the horizon problem as
\begin{eqnarray}\label{Nom-con}
d_\gamma\equiv a(t)\int\frac{dt}{a(t)}>H^{-1},
\end{eqnarray}
(where $d_\gamma$ is the particle horizon distance) associated
to an inflationary universe, which will be obtained in our herein NC model.
Therefore, we first deal about these general conditions and then we will investigate
and analyze the consequences produced by our numerical endeavors.

In the absence of the scalar potential, equation (\ref{asli3}) can also be written as
 \begin{eqnarray}\label{KG-v-zero}
\frac{d(a^3\dot{\phi})}{dt}=-a_0\kappa \theta\phi^3e^{\kappa\phi(t)}\dot{\phi},
\end{eqnarray}
where we have used (\ref{exp-rel}) and (\ref{exp-rel-2}).
Integrating (\ref{KG-v-zero}) over $dt$, we obtain
 \begin{eqnarray}\label{phidot}
\dot{\phi}=-\frac{\theta}{a_0^2 \kappa^3e^{2\kappa\phi(t)}}\left[(\kappa\phi)^3-3(\kappa\phi)^2+6\kappa\phi
-6\right]+\frac{c}{a_0^3e^{3\kappa\phi(t)}},
\end{eqnarray}
where, again, we have used (\ref{exp-rel}) and (\ref{exp-rel-2}); $c$ is an
integration constant and it equals to the initial value of $a^3\dot{\phi}$.
It is clear that $\dot{\phi}$ depends (explicitly) also linearly on the
NC parameter. By substituting $\dot{\phi}$ form (\ref{phidot}) into (\ref{asli2}), $\theta^2$
will also be present in the
relation associated to the second (time) derivative of the
scale factor, which is consistent. As mentioned, what is
important is that the NC parameter has the correct linear
dependence in the (standard form of) the equations of motion.

Employing (\ref{exp-rel}) and (\ref{exp-rel-2}) and (\ref{phidot}) in (\ref{asli2}), it is straightforward to show that
\begin{eqnarray}\label{a2dot}
\frac{\ddot{a}}{a}\!\!\!&=&\!\!\!-\frac{2c^2\kappa^2}{a_0^6 e^{6\kappa\phi(t)}}+\theta\Lambda(\phi)+\theta^2\Psi(\phi),
\end{eqnarray}
where
\begin{eqnarray}\label{lan}
\Lambda(\phi)\!\!&\equiv&\!\!\frac{3c}{a_0^5\kappa e^{5\kappa\phi(t)}}\left(\kappa\phi-2\right)\left[\kappa\phi(\kappa\phi-2)+4\right],\\\nonumber
\\
\Psi(\phi)\!\!&\equiv&\!\!-\frac{1}{a_0^4\kappa^4e^{4\kappa\phi(t)}}\left[(\kappa\phi)^2
(\kappa\phi-6)+12(\kappa\phi-1)\right]\times\left[(\kappa\phi)^2(\kappa\phi-3)+6(\kappa\phi-1)\right].\label{si}
\end{eqnarray}
From (\ref{a2dot}), we observe that the acceleration/deceleration condition of the scale factor completely
depends on the evolution of the scalar field, which, in turn, is obtained from a
nonlinear differential equation (\ref{c-zero-V-1}). Note that to
obtain $\frac{\ddot{a}}{a}$ for the commutative case, $\theta$ must be set equal
to zero in both (\ref{c-zero-V-1}) and  (\ref{a2dot}).
In what follows, when we will use the numerical analysis to get the evolution of
the scale factor, we will see how the dynamical relation (\ref{a2dot}) works.

It is worthwhile to discuss concerting a required condition pertinent to inflation.
Employing (\ref{exp-rel-2}) and (\ref{phidot}), we can easily show that
\begin{eqnarray}\label{horizon-1}
d_\gamma=\frac{a(t)}{c\kappa}\int a^2(t)H(t)dt
+\frac{\theta a(t)}{c \kappa^3}\int\left[(\kappa\phi)^3-3(\kappa\phi)^2+6\kappa\phi
-6\right]dt.\\\nonumber
\end{eqnarray}
Employing integration by parts for the first integral in the right hand side and employing (\ref{exp-rel}), we obtain
\begin{eqnarray}\label{horizon-2}
d_\gamma=\frac{a_0^3}{2c\kappa}e^{3\kappa\phi(t)}
+\frac{a_0\theta e^{\kappa\phi(t)}}{c\kappa^3 }\int\!\!\!\left[(\kappa\phi)^3-3(\kappa\phi)^2+6\kappa\phi
-6\right]dt.
\end{eqnarray}

Using relations (\ref{exp-rel-2}), (\ref{phidot}) and (\ref{horizon-2}), we get
\begin{eqnarray}\label{horizon-3}
d^{\rm nc}_\gamma\!\!\!&=&\!\!\!\frac{a_0^3e^{3\kappa\phi(t)}}{2c\kappa}+\frac{a_0\theta e^{\kappa\phi(t)}}{c\kappa^3 }\int\!\!\!\left[(\kappa\phi)^3-3(\kappa\phi)^2+6\kappa\phi
-6\right]dt
\\\nonumber
\\\nonumber
\!\!\!&+&\!\!\!\frac{a_0^3\kappa^2e^{3\kappa\phi(t)}}{a_0\theta e^{\kappa\phi(t)}\left[(\kappa\phi)^3-3(\kappa\phi)^2+6\kappa\phi
-6\right]-c\kappa^3}
\end{eqnarray}
where
\begin{eqnarray}\label{horizon-4}
d^{\rm nc}_\gamma\!\!\!&\equiv&\!\!\!d_\gamma-H^{-1},
\end{eqnarray}
To satisfy the nominal condition for an inflationary universe in our herein model, the
condition $d^{\rm nc}_\gamma>0$ must be satisfied.
In what follows, we will investigate this condition for our numerical solutions.
We should note that as $d^{\rm nc}_\gamma$ completely
depends on the scalar field, therefore, to plot the time
behavior of $d^{\rm nc}_\gamma$, we must solve the
differential equation (\ref{c-zero-V-1}). Obviously, to get $d^{\rm nc}_\gamma$ corresponds
to the commutative case, we must set $\theta=0$ in both
(\ref{c-zero-V-1}) and  (\ref{horizon-3}). We expect that the NC modifications
in (\ref{horizon-3}) may assist properly to get an appropriate nominal condition for our inflationary model.

Let us focus on a numerical analysis to depict the behavior of the above mentioned quantities.
Our numerical endeavors are summarized in what follows.

\begin{figure}
\centering\includegraphics[width=3.2in]{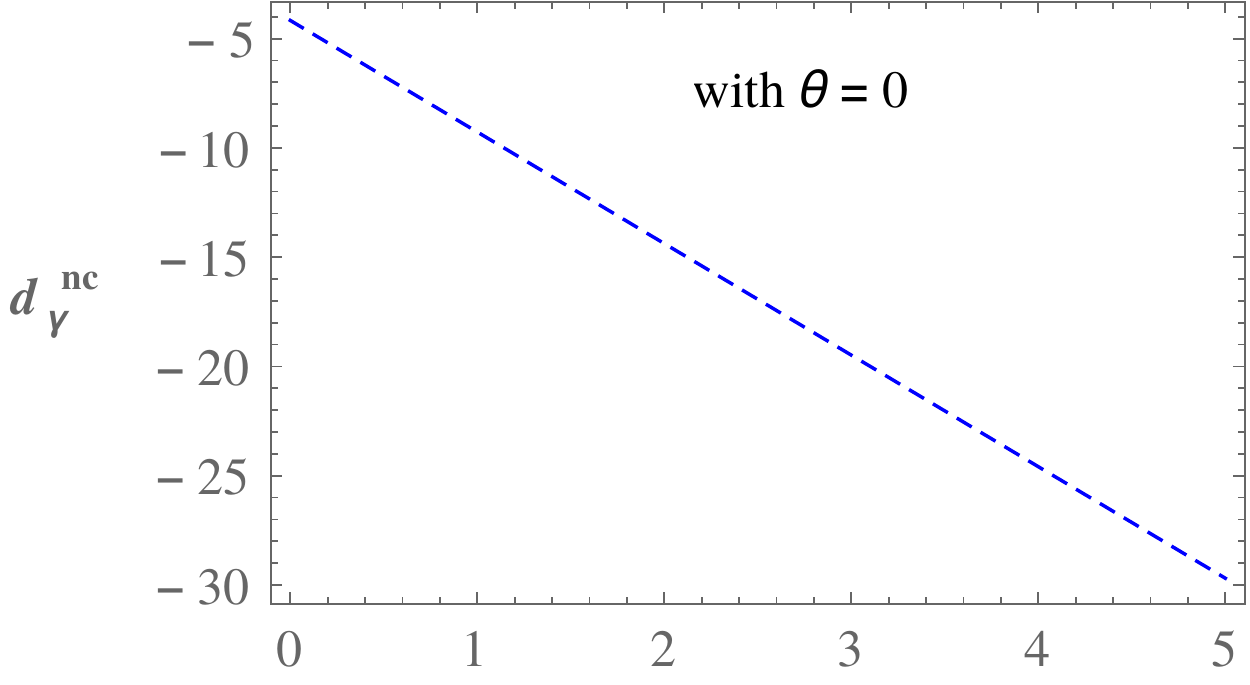}\\
\centering\includegraphics[width=3.2in]{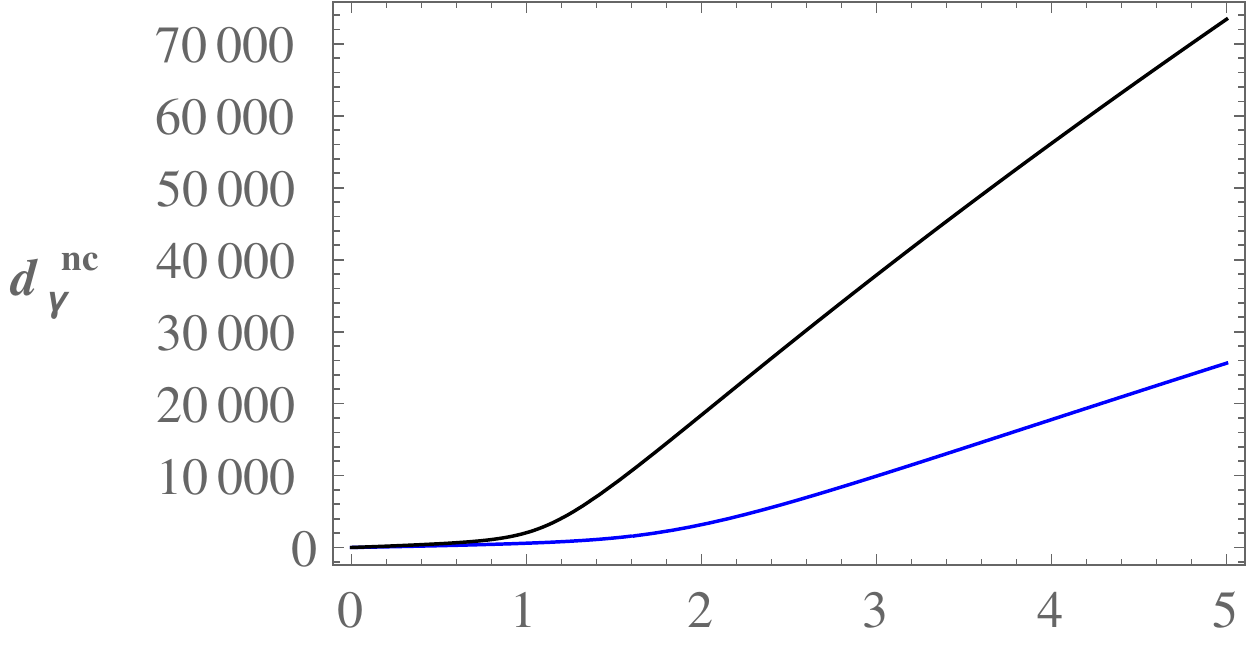}
\centering\includegraphics[width=3.2in]{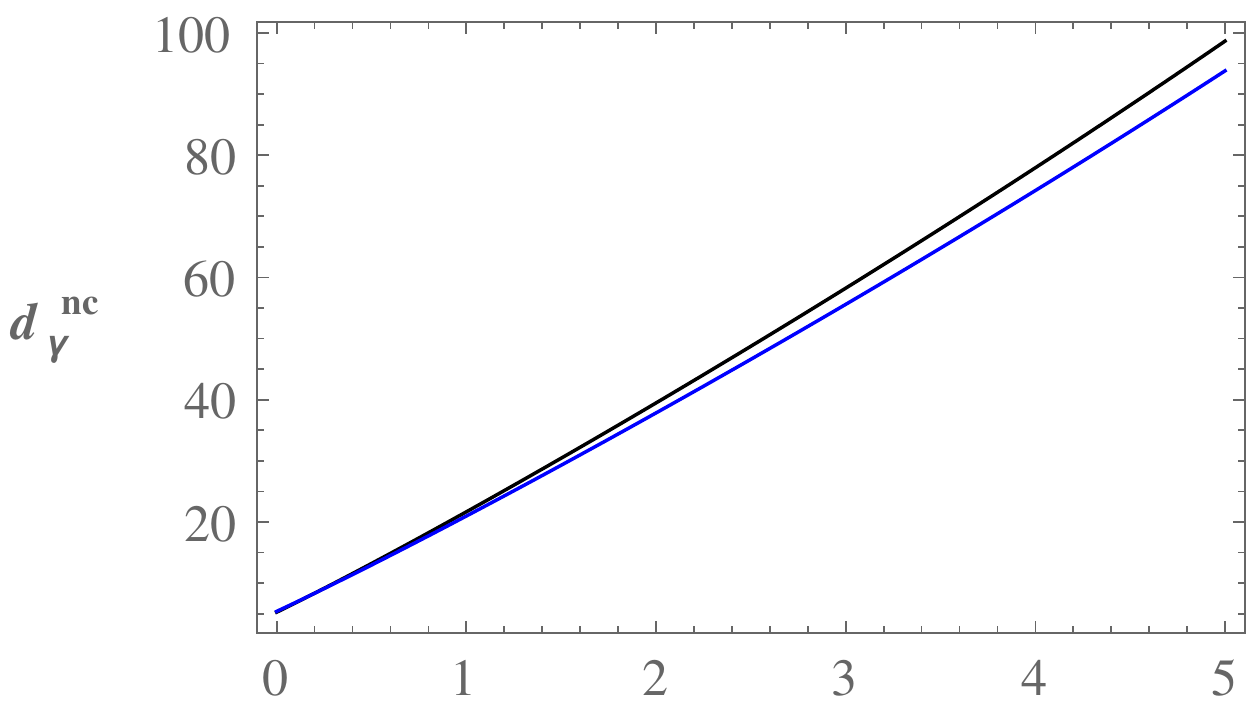}
\caption{The time behavior of $d^{\rm nc}_\gamma$ for the commutative (upper panel) and NC cases (middle and lower panels).
 In the middle panel, the black and blue curves are associated to $\theta=-0.001$ and $\theta=-0.0005$, respectively;
 While in the lower panel, they are associated to $\theta=-0.00001$ and $\theta=-0.000009$, respectively.
We have employed the other initial conditions equal to those used to plot the previous figures.
Namely, we have set $8\pi G=1$, $\kappa=\frac{\sqrt{6}}{6}$, $a_0=0.01$ ,
$\phi(0)=1=\dot{\phi}(0)$ and $c=a_0^3\dot{\phi}(0)$.}
\label{nominal}
\end{figure}

For very small negative\footnote{The small negative values of
the NC parameter, in the case where the scalar field is positive, yield
an negative $p_{\rm nc}$ which can be conjectured to drive
an accelerated expansion.} values of the NC parameter, we have observed that:
\begin{itemize}
  \item
For a small interval of the cosmic time at early times, in contrast with the commutative case,
we found an accelerating expansion for both the scale factor and scalar
field, namely, we obtained $\frac{\dot{a}}{a},\frac{\dot{\phi}}{\phi}>0$
and $\frac{\ddot{a}}{a},\frac{\ddot{\phi}}{\phi}>0$.

Subsequently, after this short
time acceleration, both
the scale factor as well as scalar field begin to decelerate
with the cosmic time, see figures \ref{phi-zero} and \ref{a-zero}.
However, for the commutative case, both of them always decelerate.
We can interpret such interesting behaviors as follows: a very short time interval in the early universe, in
which the scale factor accelerates, can be associated to a substantial
epoch of inflation. Immediately after this inflationary epoch, the scale factor
decelerates, which can be assigned to the radiation epoch.
Such an appropriate transition from an inflationary epoch to a radiation dominated era is called graceful exit.

Concretely, our NC model, contrary to the corresponding
standard scenario, even in the case of a vanishing scalar potential,
can
describe, at least qualitatively, a realistic inflationary phase for  the universe.
These  interesting consequences are associated to  NC effects, which involve  solely a NC
parameter,  which appears  linearly in the (modified) set of field equations.

\item In figures \ref{rho-p-tot-zero}, we have shown the time behavior of what we specified as the
total kinetic energy $\rho_{\rm tot}$, total pressure $p_{\rm tot}$ as well
as explicit NC pressure $p_{\rm nc}$.

We observe that $\rho_{\rm tot}$ increases during
the inflationary epoch to reach its maximum value, and subsequently, it
turns to decrease during the radiation dominated era and afterwards.
Whereas, $p_{\rm tot}$ and $p_{\rm nc}$ always get negative values, such that they decrease during the
inflationary epoch to reach their corresponding minimum values;  immediately afterwards, they  increase
during the radiation dominated era and afterwards.
Moreover, all these quantities tends to zero for large values of the cosmic time.

Let us interpret such unusual\footnote{As in the standard cosmological models, we expect
that the energy density of the universe should decrease while the cosmic time increases.} time evolution.

First, we should note that
all of these quantities  depend on the NC parameter.
Unfortunately, complicated and {\it implicit} dependence (of these quantities) on the
NC parameter does not allow us to separate analytically a strictly NC behavior.
For instance, in equation (\ref{asli2}), it seems that  $p_{\rm tot}$ is
composed of two parts, in which the NC component is separated completely as
$p_{\rm nc}=\frac{\theta\phi^{3}\dot{\phi}}{3a^2}$ from what seems, naively, the commutative part.
 However, our numerical simulation shows that, for $\theta\neq0$,
even for vanishing scalar potential, when $p_{\rm tot}-p_{\rm nc}$ is plotted against the cosmic time,
it does not correspond to the associated quantity in
the commutative model. More precisely, when $V=0$, we get $p_{\rm tot}-p_{\rm nc}=\rho_{\rm tot}$, which
is plotted as a red curve in the middle panel of figure \ref{rho-p-tot-zero}, whose time
behavior is  completely different from that is shown in the upper panel (of the same figure)
for $p_{\rm \phi}$=$\rho_{\rm \phi}$ (the pressure associated to the commutative case).
 Concretely, the pressure $p_{\rm nc}$ merely plays  an {\it explicit} NC role
 and it cannot be interpreted as the whole NC component  in $p_{\rm tot}$.
 Such  result confirms the implicit dependence on the NC parameter for those observables.
However, as expected, when we set $\theta=0$ in the numerical computation, we recover
exactly the consistent  behaviors associated to their corresponding commutative counterparts.

Secondly, by comparing the time behaviors (associated to the commutative
and NC cases) for known quantities, we observe that, for small negative values of $\theta$, during the so induced inflationary epoch the
NC effect plays its role more drastically with respect
to the radiation-dominated epoch.

Disclaimer: In order to check the degree of
accuracy for every set of numerical results, we have
depicted the numerical error in our solutions when they have
to satisfy the
conservation equation (\ref{cons}), see, e.g., figure \ref{rho-p-tot-zero}, lower panel.

\item
Up to now, we have claimed that our NC model not only yields an accelerating epoch at early times but also
such an accelerating universe, after a very short time, enters to radiation
dominated decelerating phase. However, to get a successful inflationary scenario,
we should resolve the main problem with the standard cosmology, namely the
horizon problem. In other word, we should examine our numerical results at least for the mentioned nominal condition, which we
wrote it as $d^{\rm nc}_\gamma>0$ (where $d^{\rm nc}_\gamma$ is
given by (\ref{horizon-3})). Our numerical results show that:
(i) it is never satisfied for the commutative case, as expected;
(ii) while, for the NC case, for different small (negative) values of $\theta$, by employing the same
initial conditions used to get the above described inflationary
universe, we have shown that it is satisfied during all times of the evolution of the universe, see figures \ref{nominal}.

\end{itemize}

\begin{figure}
\includegraphics[width=2.5in]{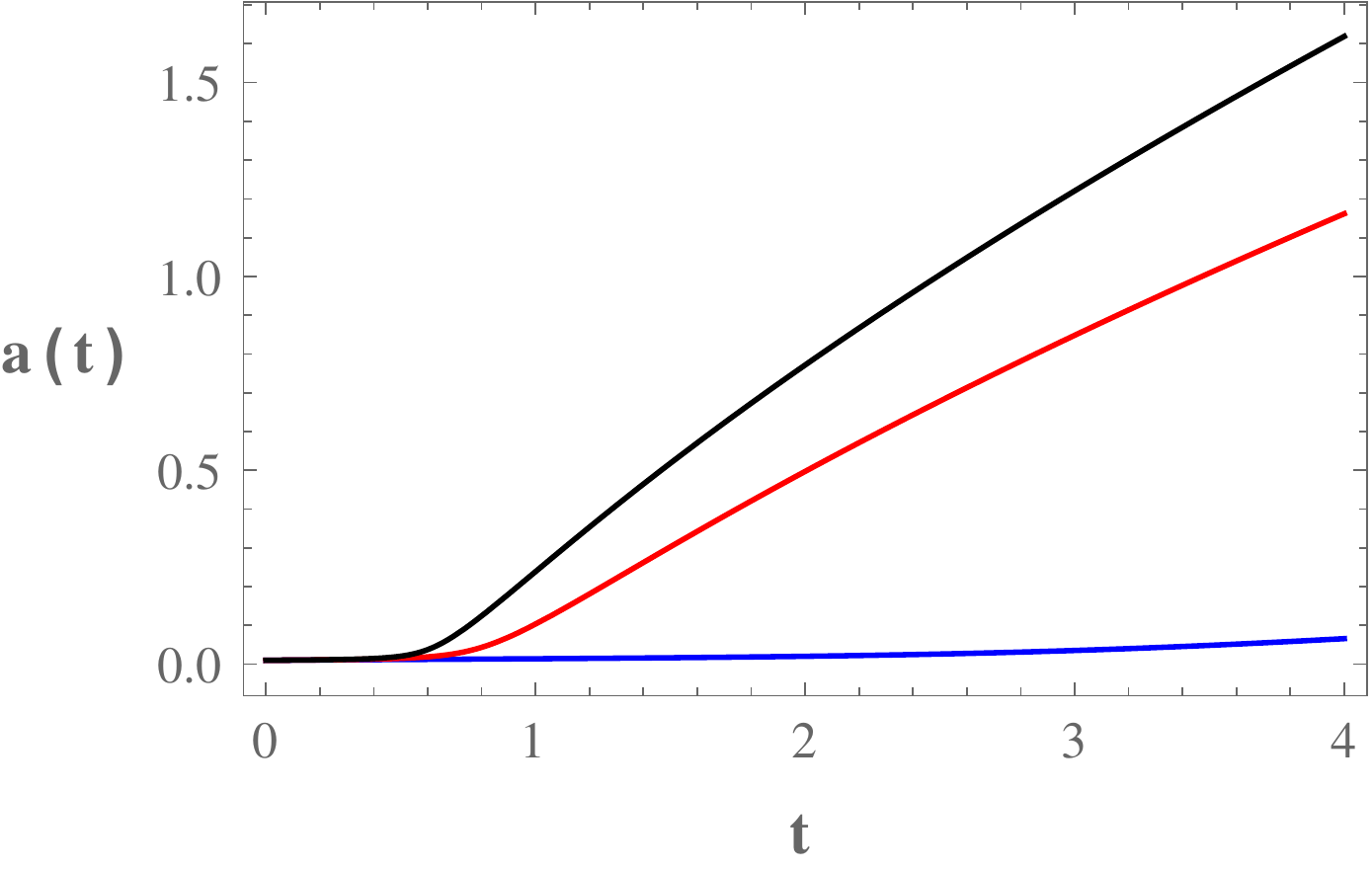}
\includegraphics[width=2.5in]{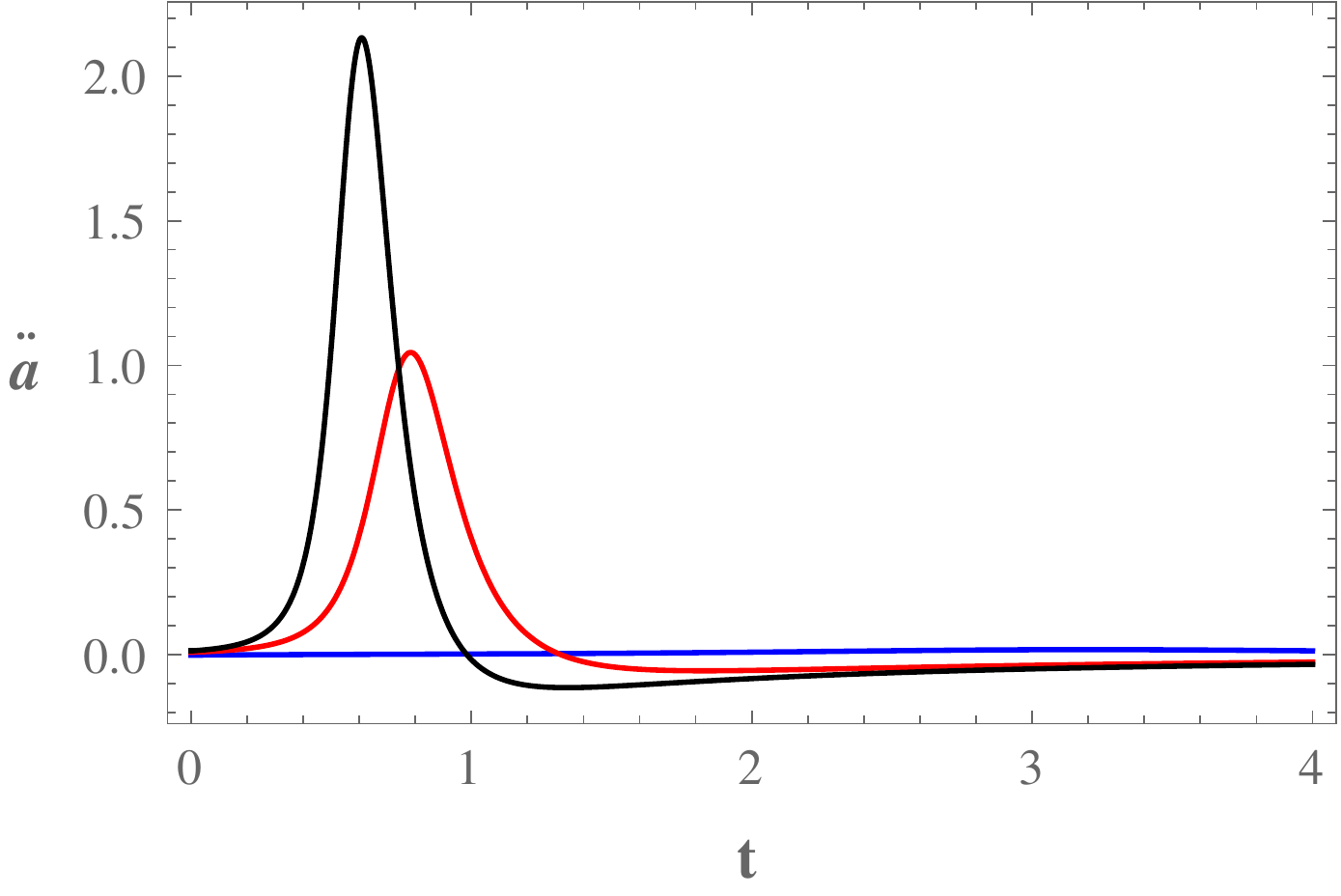}
\caption{The behavior of the scale factors (left panel) and their corresponding second time derivatives (right panel)
 against cosmic time for the NC case where $\theta=-0.00009$ (blue curves),  $\theta=-0.0007$ (red curves) and $\theta=-0.001$ (black curves).
 The black dashed line is associated to $\ddot{a}=0$ to clearly show when those quantities are positive or negative.
We have set $8\pi G=1$, $\kappa=\frac{\sqrt{6}}{6}$, $a_0=0.01$ ,
$\phi(0)=1=\dot{\phi}(0)$.}
\label{dif-value}
\end{figure}

In addition, our numerical results for the free field case, have shown that (i) choosing different values
of the NC parameter, (ii) by taking the same consistent values for the other initial conditions, then the  time
behavior of the NC quantities and their time derivatives are effectively changed.
For instance, let us consider  the behavior of scale factor against the cosmic time:
we have found that the smaller the value of $|\theta|$, the larger the time
interval and the smaller the number of e-folding associated to the
accelerating epoch (of the very early times), respectively; see, e.g., figures \ref{dif-value}.

Before closing this section, let us add a further feature concerning this important case.

It is straightforward to show that in the absence of the scalar field potential,  the evolution of the scale factor  is given by
\begin{eqnarray}\label{evol-a-1}
\frac{\dot{a}\dddot{a}}{a^2}+\left[2-\frac{3}{{\rm ln}(\frac{a}{a_0})}\right]\frac{\dot{a}^2\ddot{a}}{a^3}-\left(\frac{\ddot{a}}{a}\right)^2+
2\left[2-\frac{3}{{\rm ln}(\frac{a}{a_0})}\right]\left(\frac{\dot{a}}{a}\right)^4=0,
\end{eqnarray}%
where $a_0$ is an integration constant. Note that the NC parameter does not appear explicitly in the
field equations. However,  when $\theta=0$, instead of the above
equation, we proceed to compute  the  equation corresponding to the standard case for the free scalar field, namely,
\begin{eqnarray}\label{evol-a-2}
\frac{\dddot{a}}{a}+3\frac{\dot{a}\ddot{a}}{a^2}-4\left(\frac{\dot{a}}{a}\right)^3=0.
\end{eqnarray}

Regarding the significance of the modified evolutionary equation of
the scale factor (\ref{evol-a-1}), let us take a quick glance over
the $R^2$ (Starobinsky) inflationary model \citep{S80,V85}, in which the
Einstein equations have been solved in the presence of effective quantum corrections.
This model does have a graceful exit from an acceleration phase (associated
to the inflationary era) and it is consistent with observational
data associated to the spectrum of the primordial perturbations \citep{Planck-2015-1, Planck-2015-2}.
Briefly, this model has been considered as one of
the successful inflationary models regarding the observational constraints
imposed by the recent Planck data \citep{Planck-2015-1,Planck-2015-2}.
In this model, the evolutionary equation for the scale factor, for the
case of spatially flat FLRW line-element, is given by \citep{S80,V85}
\begin{eqnarray}\label{evol-a-3}
2\frac{\dot{a}\dddot{a}}{a^2}+2\frac{\dot{a}^2\ddot{a}}{a^3}-\left(\frac{\ddot{a}}{a}\right)^2
-\left(\frac{M_0^2}{H_0^2}+3\right)\left(\frac{\dot{a}}{a}\right)^4+M^2\left(\frac{\dot{a}}{a}\right)^2=0,
\end{eqnarray}
where $H_0^2=360\pi/Gk_2 $ and $M_0^2=-360\pi/Gk_3$, where
 $k_2>0$ and $k_3<0$ are numerical coefficients.

 It is not feasible  to establish a full matching  correspondence between our herein
 NC framework and  $R+R^2$ model, specifically at the Lagrangian
 level.
 Nevertheless, at the level of the field
 equations (\ref{evol-a-1}) and (\ref{evol-a-3}) [or their corresponding
 phase space plane, by assuming appropriate approximations with respect to the corresponding epochs],
 we can speculate in extracting the following. As the integration constant $a_0$ relates the scale factor
 and the scalar field (which, in turn, is obtained from Klein-Gordon
 equation and it depends on the NC parameter) via relation (\ref{exp-rel}) for a
 constant time, it may be possible to consider associating  the NC parameter to the coefficients $k_2$ and $k_3$.

\section{Noncommutative Setting and Slow-Roll Approximations}
\label{sec-SRA}

Let us now in what follows, discuss the case where a potential is present.

The SRA~\citep{SL93}, which leads to reliable consequences when employing
 smooth potentials, have been
traditionally applied as an approximation method in inflationary
cosmology. However, there are also other alternative
approximations, such as the WKB~\citep{MS03}, the Green
function~\citep{GS01} and the improved WKB~\citep{CF05} methods,
that have been used by some researchers in the study of inflationary scenarios.
Nevertheless, in this work, we employ the SRA for our herein
NC model and then analyze the results according to this context.

In the commutative case, in order to attain an inflationary
accelerating universe, the potential energy of inflation must
dominate the kinetic energy of the system. Consequently, to get a
sufficient amount of inflation, a flat potential associated to an
inflationary scenario is needed. Concretely, it is required to impose
the slow-roll conditions
\begin{equation}\label{SRA}
\frac{1}{2}\dot\phi^2\ll V(\phi) \qquad {\rm and}\qquad
\ddot{\phi}\ll 3H\dot{\phi},
\end{equation}
with the slow-roll parameters being defined as
\begin{equation}\label{9}
\epsilon \equiv -\frac{\dot H}{H^2} \qquad {\rm and}\qquad \eta
\equiv -\frac{|\ddot{\phi}|}{|\dot{\phi}|H},
\end{equation}
where, in the case of $\theta =0$, using (\ref{dh}), we get obviously
$\eta = -|\ddot{H}|/(2|\dot{H}|H)$. In this case, from conditions
(\ref{SRA}), it is straightforward to show that, during inflation,
these parameters should satisfy $\epsilon\ll1$ and
$|\eta|\ll1$~\citep{wein2}; at the end of inflationary phase,
these parameters increase with the order of unit and thus, the
SRA breaks down.

However, for the NC configuration, in order to
satisfy the condition $\epsilon\ll1$,
 from equations
(\ref{asli1}) and (\ref{dh}), we get the  condition
\begin{equation}
\dot\phi^2\ll |V(\phi)-\frac{1}{2}\theta a^{-2}\phi^{3}\dot{\phi}|.
\end{equation}

Now, by imposing these conditions, the most slowly varying terms
in equations of motion (\ref{asli1}), (\ref{asli2}) and
(\ref{asli3}) become negligible and hence, it remains
\begin{eqnarray}\label{SREq1}
H^{2}\approx \frac{8\pi G}{3}V(\phi),
\end{eqnarray}
\begin{eqnarray}\label{SREq2}
2\frac{\ddot{a}}{a}+H^2\approx 8\pi G\left[V(\phi)-\frac{\theta\phi^{3}\dot{\phi}}{3a^{2}} \right],
\end{eqnarray}
\begin{equation}\label{SREq3}
3H\dot{\phi}+V'(\phi)+\theta\dot{a}\left(\frac{\phi}{a}\right)^{3}\approx
0.
\end{equation}
Likewise, by employing Eqs.~(\ref{SREq1}) and (\ref{SREq3}), the
time derivative of the Hubble parameter, according to
Eq.~(\ref{dh}), reduces to
\begin{equation}\label{Hdot}
\dot{H}\approx -\frac{4\pi G}{3}\left[\frac{V'^{\, 2}(\phi)}{V(\phi)}+\theta
\frac{\sqrt{3V(\phi)}V'(\phi)}{3V(\phi)}\frac{\phi^{3}}{a^{2}}\right].
\end{equation}

Hence, the slow-roll parameters in (\ref{9}), will read now as
\begin{equation}\label{16}
\epsilon \approx \epsilon_{1}+\theta
\frac{\sqrt{3V(\phi)}V'(\phi)}{6V^{2}(\phi)}\frac{\phi ^{3}}{a^2}
\end{equation}
and
\begin{equation}\label{ett}
\eta\!\approx\!\eta_{1}\!\!-\!\epsilon\!+\!\theta
\sqrt{3V(\phi)}\frac{\phi^{3}}{a^{2}}\!\left(\!\frac{1}{\phi
V(\phi)}\!\!+\!\!\frac{2+\epsilon}{3V'(\phi)\!\!+\!\!\theta\sqrt{3V(\phi)}
\frac{\phi^{3}}{a^{2}}}\!\right)\!\! ,
\end{equation}
where, analogous to the standard
model~\citep{Tasi}, we have
defined $\epsilon_{1}$ and $\eta_{1}$ as
\begin{equation}\label{e1}
\epsilon_{1}\equiv\frac{V'^{\, 2}(\phi)}{2V^{2}(\phi)}
\end{equation}
and
\begin{equation}\label{eta1}
\eta_{1}\equiv \frac{V''(\phi)}{V(\phi)}.
\end{equation}
Moreover, note that relation~(\ref{16}) is obtained by employing
equations~(\ref{SREq1}) and (\ref{Hdot}) in the first definition
of (\ref{9}), and relation (\ref{ett}) has been derived by taking
the time derivative of equation (\ref{SREq3}), then employing
equations (\ref{SREq1}) and (\ref{SREq3}) into the second
definition of (\ref{9}). It is clear that by setting $\theta= 0$
in (\ref{16}) and (\ref{ett}), the NC slow-roll parameters reduce to
their corresponding commutative ones.


\subsection{Behaviors of physical quantities in the presence of the polynomial scalar field potential}
\label{numerical}

Let us therefore investigate in this section the  behaviors of cosmological quantities,
with the aid of a numerical analysis,  while
considering two particular cases of the following
general scalar potential
\begin{equation}
V(\phi)=\lambda M^{4}\left(M/\phi\right)^{n},\label{5}
\end{equation}
where $\lambda>0$ is a parameter, $M$ is some mass scale and $n$
is a positive or negative integer constant. Potential~(\ref{5})
has been studied in the context of chameleon field
theory~\citep{weltman1,gubser,saba}; in
particular, when $\lambda=1$ and $n >0$, it is called the
Ratra–Peebles potential,
that is used in the intermediate inflation~\citep{B90,M02} and in
the quintessence models. Furthermore, when $n\neq-4$, $M$ can be
scaled such that, without loss of generality, we can set
$\lambda$ equals to unity~\citep{weltman1}. Whereas for
$n=-4$, $M$ drops out and the $\phi^{4}$ theory is
resulted~\citep{gubser}.
Also note that, action~(\ref{eq1}) with a
scalar potential $V(\phi)$ can be considered as the Einstein
representation of the well-known Brans-Dicke theory whose
corresponding Jordan frame exists with a trapped field and a
coupling function $\omega(\phi)$~\citep{GL96}.

 In this section, polynomial chaotic inflation, in which the scalar
 potentials are given by $V(\phi)=M^2\phi^{2}$ (massive scalar field)
 and $V(\phi)=\lambda\phi^{4}$ (self-interacting scalar field) constitute the focus of our interest.
  It has been established that, for  standard models, with the
  mentioned scalar potentials, inflation occurs while the scalar
field rolls down towards the potential minimum.

By substituting potential~(\ref{5}) into equations (\ref{asli1}),
(\ref{asli2}) and (\ref{asli3}), we obtain
\begin{equation}
H^{2}=\frac{8\pi G}{3}\left(\frac{1}{2}\dot{\phi}^{2}+\frac{\lambda M^{4+n}}{\phi^{n}}\right),\label{6}
\end{equation}
\begin{equation}
2\frac{\ddot{a}}{a}+H^{2}=
-8\pi G\left[\left(\frac{1}{2}\dot{\phi}^{2}-\frac{\lambda M^{4+n}}{\phi^{n}}\right)+\frac{\theta\phi^{3}\dot{\phi}}{3a^2}\right],\label{dps-eq2}
\end{equation}
\begin{equation}
\ddot{\phi}+3H\dot{\phi}-\frac{n\lambda M^{4+n}}{\phi^{n+1}}+
\theta\dot{a}\left(\frac{\phi}{a}\right)^{3}=0.\label{dps-eq3}
\end{equation}
It is easy to show that these equations can be rewritten as
\begin{equation}
\left(3-4\pi G{\overset{\ast}{\phi}}^2\right)H^{2}=\frac{8\pi G\lambda M^{4+n}}{\phi^{n}},
\end{equation}
\begin{equation}
2H{\overset{\ast}{H}}+\left(3+4\pi G{\overset{\ast}{\phi}}^{2}\right)H^2=8\pi G\left(\dfrac{\lambda M^{4+n}}{\phi^{n}}-\dfrac{\theta\phi^{3}{\overset{\ast}{\phi}}}{3e^{2N}}\,\right),
\end{equation}
\begin{equation}\label{friction}
{\overset{\ast\ast}{\phi}}+\left(3+\dfrac{{\overset{\ast}{H}}}{H}\right){\overset{\ast}{\phi}}-\frac{n\,
\lambda M^{4+n}}{H^{2}\,\phi^{n+1}}+\dfrac{\theta\phi^{3}}{He^{2N}}\,=0,
\end{equation}
where the asterisk $\ast$ denotes the derivative with respect to
the logarithmic scale factor $N=\ln\, a$. In this setting, the
slow-roll parameters can be rewritten as
\begin{equation}
\epsilon\equiv -\frac{\dot
H}{H^2}=-\dfrac{{\overset{\ast}{H}}}{H}\quad{\rm
and}\quad\eta\equiv
-\frac{|\ddot{H}|}{2|\dot{H}|H}=\dfrac{{\overset{\ast}{\epsilon}}}{2\epsilon}-\epsilon.
\end{equation}

In what follows, by means of numerical methods, we investigate
the behaviors of the cosmological quantities such as the
slow-roll parameter $\epsilon$ and the Hubble parameter
 within the framework of deformed phase space. Then, we compare them with their corresponding counterparts in the
commutative case.
The SRA setting is taken as a
method to obtain the analytical interpretation of these cosmological quantities.

It is important to note that, as the results of our
numerical endeavors show, the last term in (\ref{friction}), which includes the
NC parameter, behaves like an extra friction (or antifriction) term in classical mechanics.
 Moreover, we should note that among those three equations of motion, only two of
 them are independent. Hence, in order to solve them numerically, we will
 employ the Friedmann equation for the consistency of initial conditions
as well as the consistency check of the integration routine;
we consider the other two as dynamical field equations.

\begin{figure}[h!]
\includegraphics[width=2.5in]{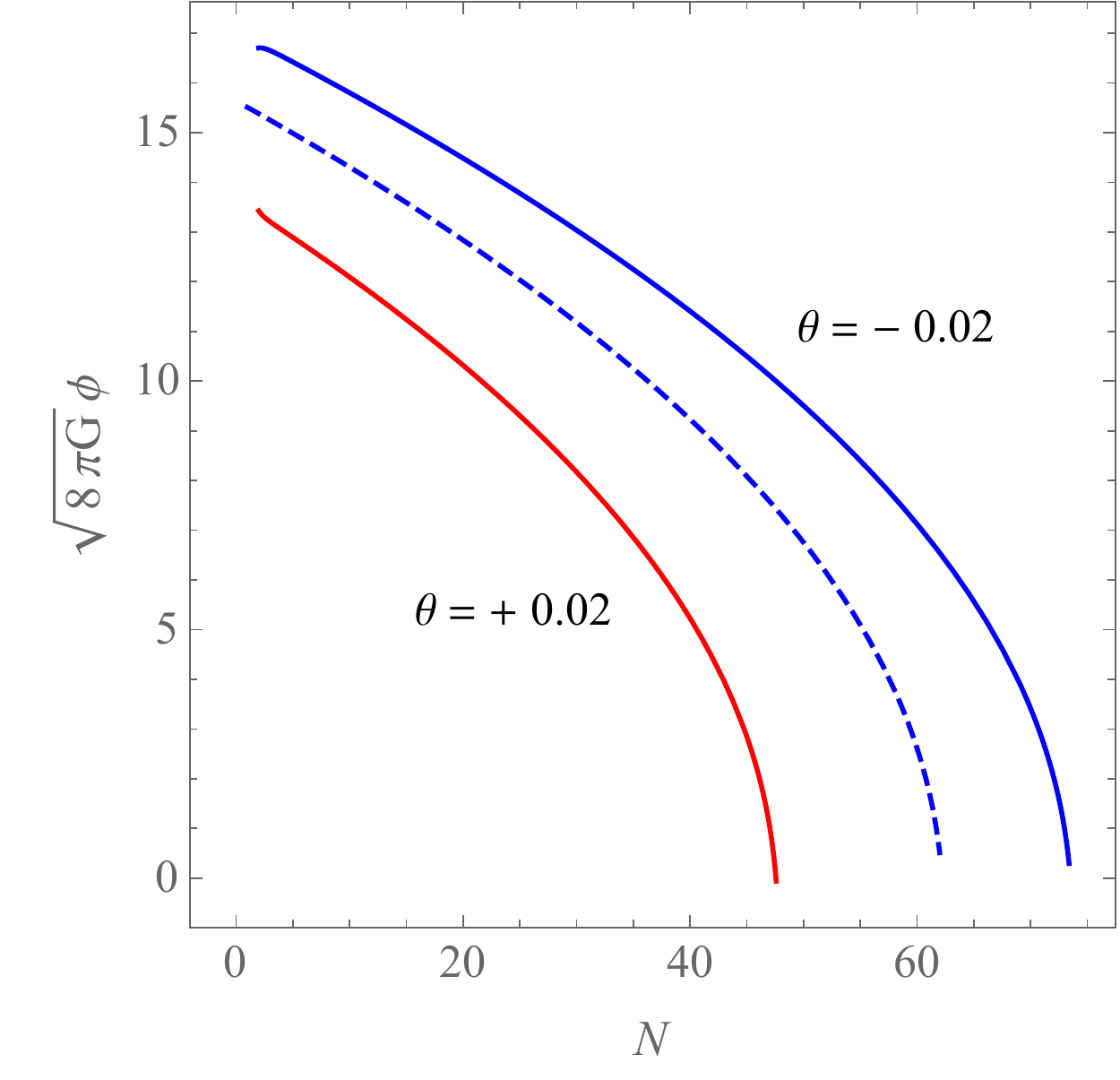}
\includegraphics[width=2.5in]{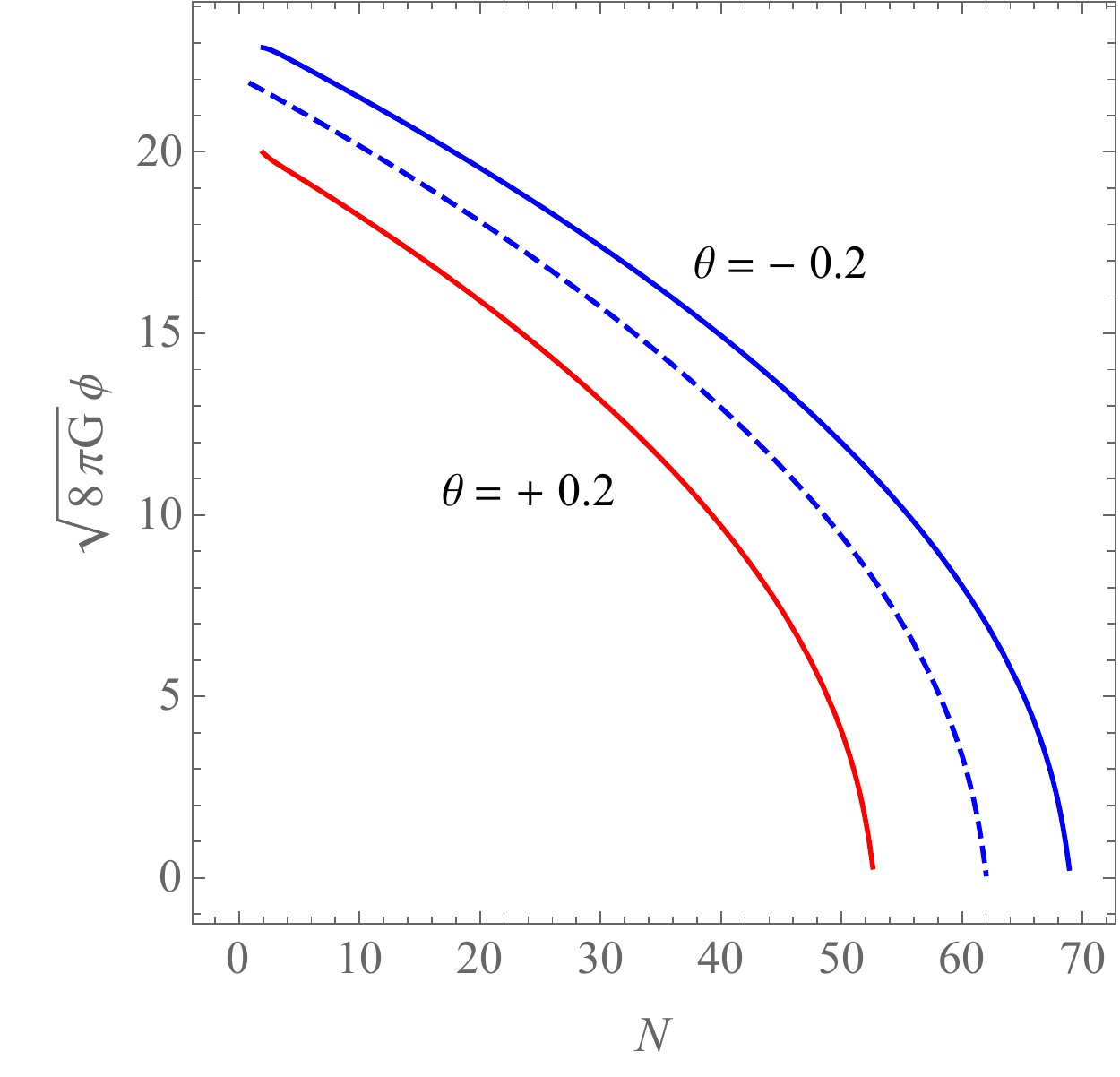}
\caption{We present a numerical evaluation of
 the inflaton scalar field $\phi$ versus $N$ for the commutative
(dashed line) and NC (solid line) cases,
where $\phi_{i}\simeq 15.6$, $\dot{\phi_{i}}\simeq -0.11$ when $n=-2$ (left panel)
whereas $\phi_{i}\simeq 22$, $\dot{\phi_{i}}\simeq -0.11$ when $n=-4$ (right panel).}
\label{Field}
\end{figure}

\begin{figure}[h!]
\centering\includegraphics[width=2.2in]{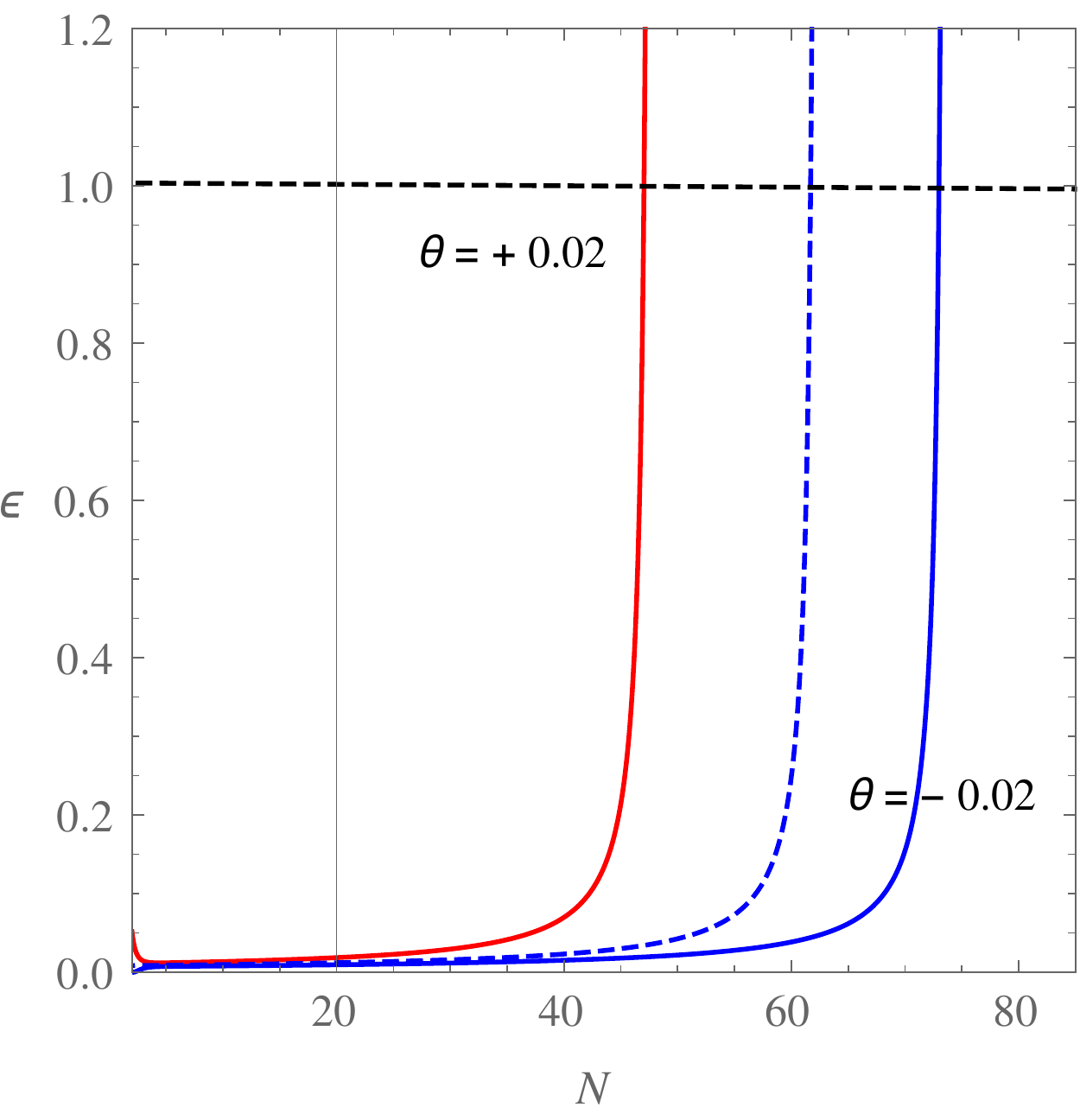}\\
\centering\includegraphics[width=2.2in]{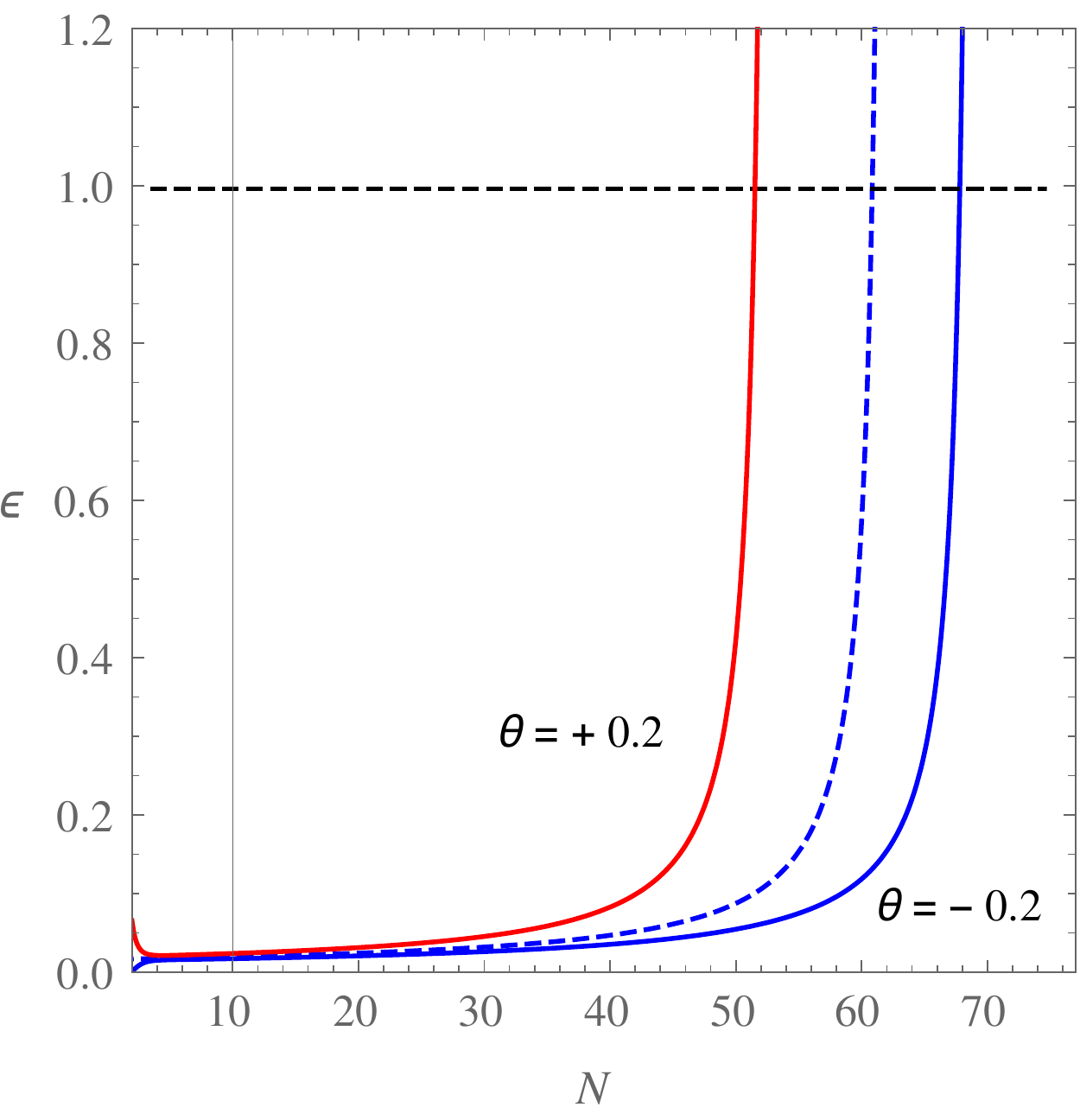}
\caption{We present a numerical evaluation of
 $\epsilon$ versus $N$ for the commutative
(dashed line) and NC (solid line) cases,
where $\phi_{i}\simeq 15.6$, $\dot{\phi_{i}}\simeq -0.11$ when $n=-2$ (upper panel)
whereas $\phi_{i}\simeq 22$, $\dot{\phi_{i}}\simeq -0.11$ when $n=-4$ (lower panel).
The slow roll parameter $\epsilon$ goes to 1 when inflation ends.}
\label{epsN1}
\end{figure}

\begin{figure}[h!]
\includegraphics[width=2.2in]{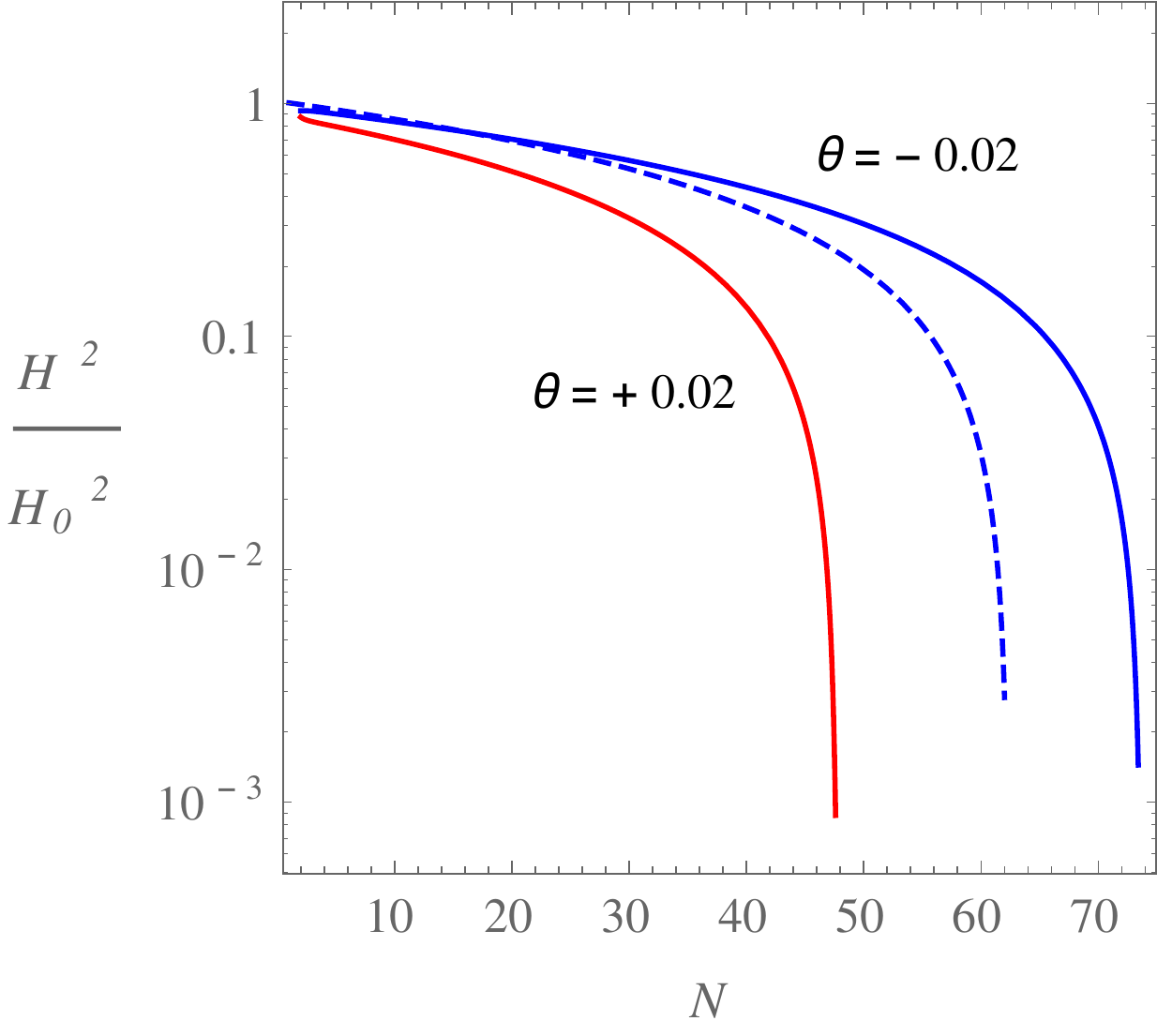}
\includegraphics[width=2.2in]{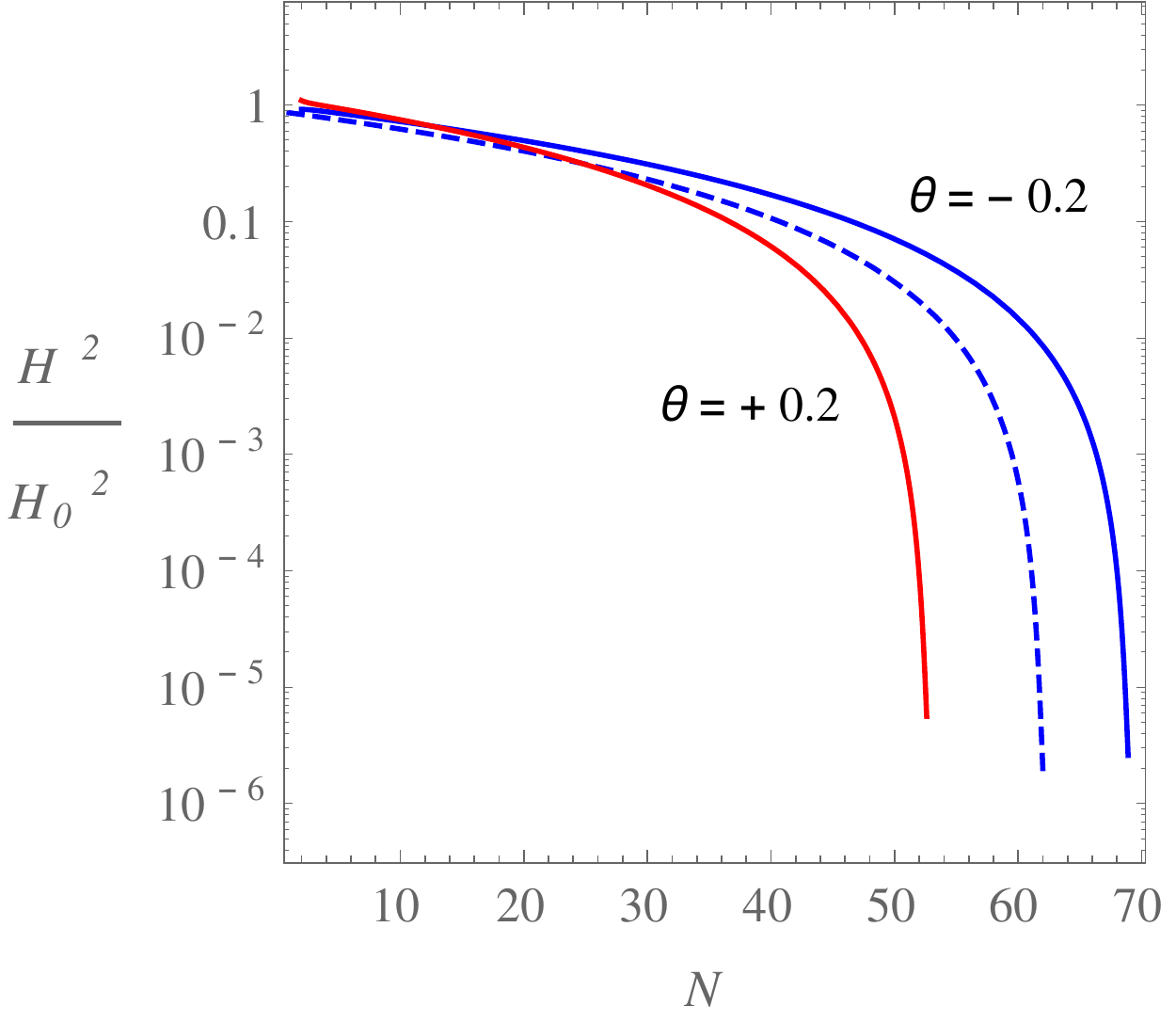}
\caption{We present a numerical evaluation of the Hubble parameter $H$ versus $N$ for the
commutative (dashed line) and NC (solid line) cases,
where $\phi_{i}\simeq 15.6$, $\dot{\phi_{i}}\simeq -0.11$ when $n=-2$ (left panel)
whereas $\phi_{i}\simeq 22$, $\dot{\phi_{i}}\simeq -0.11$ when $n=-4$ (right panel). The Hubble parameter
has been rescaled to its initial value, i.e.  $H_0$,  when inflation starts.}
\label{HubDN1}
\end{figure}

Figs.~\ref{Field} to \ref{HubDN1} show the behaviors of the
inflation scalar field, slow-roll parameter $\epsilon$ and the Hubble parameter associated
to the commutative and NC cases as a function of the e-folding number $N$, for $n=-4$ and $n=-2$ as
typical examples for the potential~(\ref{5}). Based on our numerical graphs, we observe that,
for a set of suitable initial conditions\rlap,\footnote{Following the footnote 3, the figures have been plotted by taking a new dimensionless scalar
field as $\varphi=\sqrt{8\pi G}\phi$. In such units, we have taken the NC
parameter from the interval $-1<\theta<1$.} which (except $\theta$) are
the same for both the commutative and NC cases, the
inflationary scenario associated to the commutative
case is retrieved with the correct number of e-folding
$N\approx 60$. Whilst, for the NC
counterpart, the number of e-folding either increases or decreases.
In fact, this behavior can be expected upon a close inspection of Eq.~(\ref{16}).
Taking a negative (positive) value for the NC parameter $\theta$ increases
(decreases) the number of e-folding with respect to the commutative case.

\subsection{Noncommutative case with the exponential scalar field potential}
\label{numerical}

\begin{figure}[h!]
\centering\includegraphics[width=2.2in]{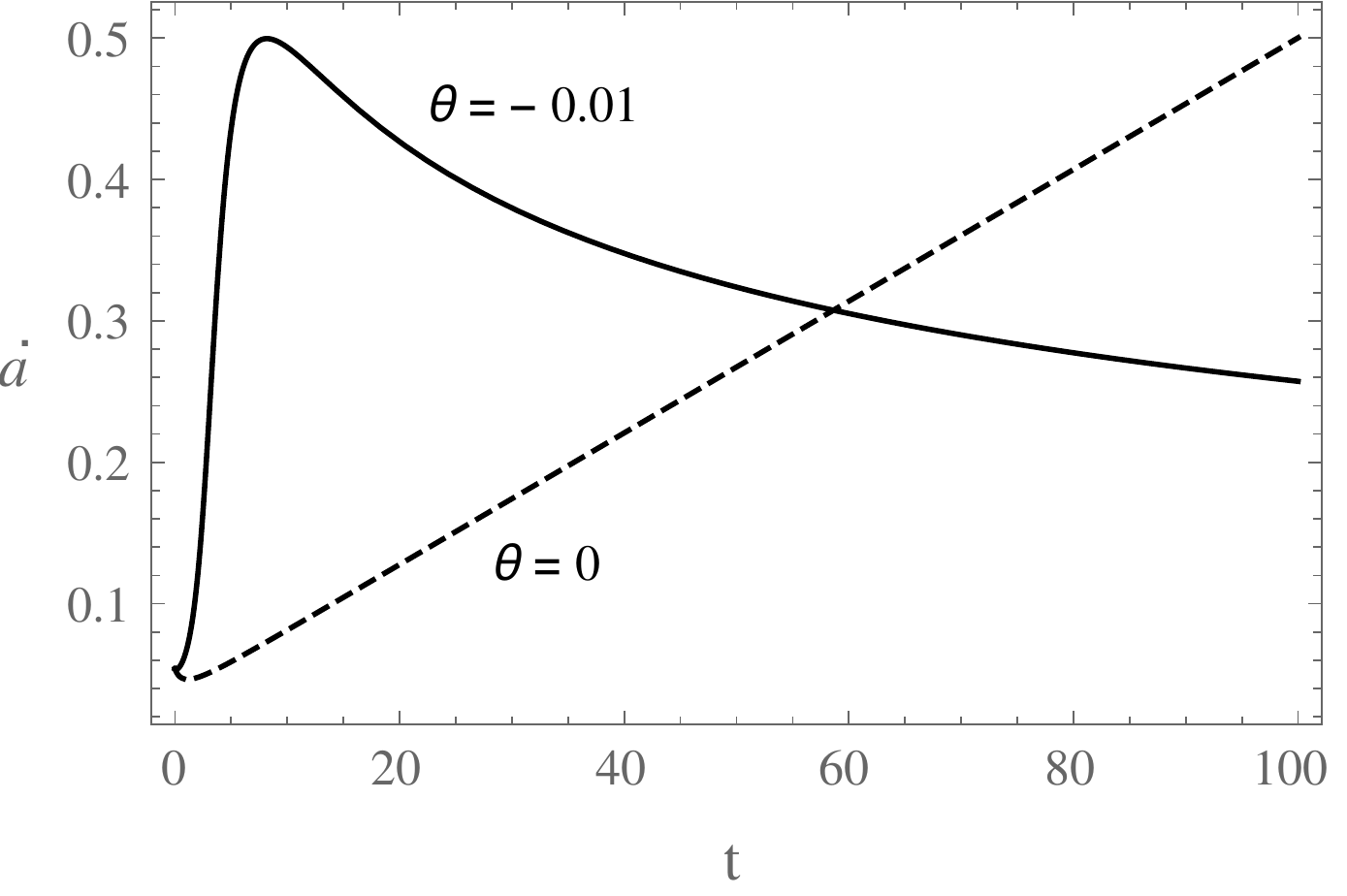}
\caption{The behavior of the scale factor first derivative,
against cosmic time, for a scalar field under the influence of an
exponential potential \eqref{exp-pot}. In the NC case (full line),
the accelerated and decelerated phases are present and smoothly connected.
In the commutative case (dashed line) we have the traditional accelerated phase
of a PLI regime.
We have set $8\pi G=1$, $\kappa=\frac{\sqrt{6}}{6}$, $a_0=0.01$ ,
$\phi(0)=1=\dot{\phi}(0)$.}
\label{exppot-f1}
\end{figure}

It has been shown that the case of the canonical scalar field model with an
exponential potential,
\begin{equation}
V\left( \phi \right)= V_0 e^{-\tilde{\kappa} \phi}\quad (\tilde{\kappa}^2=8\pi G)\: ,
\label{exp-pot}
\end{equation}
yields power law inflation (PLI)~\citep{LM85,JH87,YM88}.
It has been also extensively known that there are
two important problems with canonical PLI scenario~\citep{US13}:
(i) The range of the tensor-to-scalar ratio $r$ predicted in
these models is well above the limit reported by the Planck data.
(ii) These models suffer from the graceful exit problem.
In the scope of the present work we only address to the second problem,
namely studying numerically the evolution of the scale factor when a small negative
NC parameter $\theta$ is switched on. In Fig.~\ref{exppot-f1} we have depicted
the time evolution of the first derivative of the
scale factor for the NC and commutative cases.
As expected, in the commutative case the numerical simulation
shows an accelerated expansion of the universe in a regime according to
the well known power law evolution of the scale factor. In contrast,
when a small negative $\theta$ is taken in Eq.~\eqref{asli2},
and for the same initial conditions taken for the commutative case,
the scale factor derivative evolves in a transition between
an initial accelerated phase and a subsequent deceleration.
Therefore, the same scenario is emerging, as before, when we assumed the
free field case or with the scalar field under the influence of a polynomial potential.

In face of the overall qualitative behavior
of the evolution of the scale factor in the NC context
studied here, we can assume that the smooth transition between a
period of accelerated and a decelerated phase of expansion of the universe
was induced by means of the NC effects. Moreover, this approach has the potential
benefit of providing an alternative way of dealing with the graceful exit problem
that weakens well motivated inflationary scenarios.

\section{Conclusions}

It has been proposed that a deformation in the phase space
structure can be considered as an appropriate approach   (i) with which to discuss quantum gravity effects
and (ii) from which to predict  cosmological phenomena at very small scales.

In this work, we assumed the spatially flat FLRW line element
as the background geometry and the well know Lagrangian in
which the gravity and the scalar field are minimally coupled.
Then, we proposed a particular type of dynamical deformation
for the canonical momenta of the scale
factor and of the scalar field.
The main motivations for this choice are: (i) the simplicity
of an extra term linear in the NC parameter $\theta$ affecting the
Friedmann acceleration equation as well as the Klein-Gordon equation;
(ii) the Friedmann (Hamiltonian) constraint remains unaffected.
These last two aspects enable us, in contrast
to what happens in \citep{GSS11}, to access NC effects for the free field case and, therefore,
to build a kinetic inflation scenario. We should note that the interesting dynamics produced by our
NC choice can describe leastwise the early universe more appropriate than that
provided by other possible noncommutativity, which can be proposed between the present variables.
  Moreover, we should note that all the consequences produced by this NC idea
are entirely new and have not been presented elsewhere.

Using the Hamiltonian
formalism, we have obtained the NC equations of motion, which
are reduced to the corresponding ones in the
commutative case, when the NC parameter goes to zero.

As explained in section \ref{free potential}, our model, which bears just a one (linear) NC parameter, can
generate a suitable interesting inflationary scenario with a graceful exit, in the
absence of a scalar field potential.
Moreover, in this inflationary epoch, we
have shown that the relevant nominal condition is
perfectly satisfied during evolution of the universe at all times.
We should note that our (numerical) results have been obtained by
taking very small values of the NC parameter. Moreover, when the NC
parameter vanishes, we recover all the corresponding results associated to the standard
commutative case.
For the latter regime, in the absence
of the scalar potential, we have shown, analytically and numerically, that there is no accelerating epoch for
the universe, nor is there any satisfaction for the nominal condition.
Notwithstanding the previously stated, the NC case
provides an interesting element for analysis. In fact, from the modified evolutionary equation of
the scale factor (\ref{evol-a-1}), we can, for the sake of discussion, (formally) consider
the $R^2$ (Starobinsky) inflationary model \citep{S80}, contrast
 equations (\ref{evol-a-1}) and (\ref{evol-a-3}): as the integration constant $a_0$ relates the scale factor
 and the scalar field (which, in turn, is obtained from Klein-Gordon
 equation and it depends on the NC parameter) via relation (\ref{exp-rel}) for a
 constant time, we may speculate to relate the NC parameter to the coefficients $k_2$ and $k_3$.

By employing the SRA procedure, we have retrieved
  the approximation conditions for our herein NC setting when a potential is present for the scalar field.
 These relations can be considered as the
 generalized versions of those in the standard commutative setting, such that
 when the (constant) NC parameter tends to zero,
 all the slow-roll relations/parameters are reduced to
 those introduced in the commutative standard case.

Subsequently, by assuming a typical potential,
we have rewritten the NC equations in
terms of the scale factor as well as the logarithmic scale factor.
More concretely,
we have considered the polynomial chaotic inflation, in which the scalar
 potentials are given by $V(\phi)=M^2\phi^{2}$ (massive scalar field)
 and $V(\phi)=\lambda\phi^{4}$ (self-interacting scalar field).

 By choosing a few sets of suitable initial conditions and working in a
 re-scaled units in which the NC parameter can be of
 order unity\footnote{For instance, we have taken two values of the allowed
 NC parameter to plot the figures.}, we explored, numerically,
the behavior of the scalar field, slow-roll parameter $\epsilon$ and the Hubble parameter during the
inflation (associated to the commutative and NC
cases) against the e-folding number ${\rm N}$.
With the same initial conditions (except for $\theta$) for
both the commutative and NC cases, our numerical
results have shown that, in the commutative case, we can obtain
an inflationary universe in which the number of e-folding takes value $60$.
However, for the NC counterpart, the number
of e-folding either increases or decreases.

In what follows, it is worthwhile to mention a few points
regarding the strengths
as well as shortcomings  of our
herein model, which should be compared with the other (classical)
NC scenarios in the literature:

\begin{itemize}
\item
 In our model, all the NC field
equations were modified through a sole linear function of the
NC parameter, such that they reduce to those in
the standard case when $\theta$ goes to zero.

\item We have seen that the explicit NC corrections are weighted
as products $\theta(\frac{\phi^3\dot{\phi}}{3a^2})$
and $\theta(\frac{\phi^3\dot{a}}{a^3})$ in
equations (\ref{asli2}) and (\ref{asli3}), respectively, while the Friedmann equation is not modified.
Specifically, not only these corrections appear as  different modifications
in the field equations, but they can also influence the cosmological evolution in
the early inflationary stage (when $a(t)$ is small) as well as late times
(where $\dot{a}$ could be very large,
although in this case a strong suppression of the NC term is due to the $a^3(t)$ dependence).
We should mention that the Poisson bracket deformation
described in Eq. \eqref{deformed} can be taken as a limit for small $a(t)$,
suitable to study early time cosmology and that for large values of $a(t)$
another appropriate limit could provide NC effects for the late time cosmology.

Consequently, we claim that our model can be an appropriate
model to investigate the inflationary scenario as well as late time accelerating universe.


\item
Moreover, we should notice a possible (formal) resemblance between  the consequences obtained numerically (in the case of the free scalar potential) for evolution of the scale factor in our herein NC model and the corresponding ones associated to the kinetic inflation in Brans-Dicke theory in the presence of a different choice of noncommutativity~\citep{RM14}.

\item One of the most important shortcomings with the standard PLI models is the graceful exit problem.
We have numerically shown that, by employing the same initial conditions as used in
the commutative case, this problem is solved, appropriately, in our NC model.
Concretely, we have shown that the short time accelerating scale factor
in the presence of the exponential scalar field does connect smoothly to a decelerating epoch.
 \end{itemize}



 Finally, concerning the shortcoming of the present model, let us point following. One of the most important achievements
 of the standard inflationary scenarios is predicting the
(quantum) fluctuations behavior.
In fact, within just a few remaining concrete  inflationary scenarios (studied
within  standard commutative settings), still
viable after the Planck 2015 data survey \citep{Planck-2015-1,Planck-2015-2}, there
were crucial observable quantities, such as the scalar and tensor
power spectrum as well as the scale invariant
spectral index, which demonstrated a very good agreement with the
 observational data.

For our herein model, investigating the fluctuations and their dynamics
is a
very meaningful and significant issue, for not only to compare with the
observational data but also, to study the stability and viability of the model.
Moreover, investigating the effects of  inhomogeneous arbitrary initial conditions for
late time behaviors  would  also be a substantial outlook to proceed  from the herein NC model.
However, undertaking such significant questions
requires an evident amount of complicated calculations, to compute perturbations
for our herein NC model;  this has been left  out of the scope of the present work and
it would be studied in subsequent works.

\section*{Acknowledgments}

SMM Rasouli appreciates for the support of grant
SFRH/BPD/82479/2011 by the Portuguese Agency Funda\c{c}\~ao para a
Ci\^encia e Tecnologia. The research is supported by the grant UID/MAT/00212/2013
and COST Action CA15117 (CANTATA).

%

\begin{thebibliography}{00}

\bibitem[Snyder(1947)]{SNY}
H.S. Snyder, ``Quantized space--time", {\it Phys. Rev.} {\bf 71}, 38 (1947).
\bibitem[Snyder(1947)]{SNY1}
H.S. Snyder, ``The electromagnetic field in quantized space-time", {\it Phys. Rev.} {\bf 72}, 68 (1947).
\bibitem[Douglas(2001)]{noncom1} M.R. Douglas and N.A. Nekrasov, ``Noncommutative field theory", {\it Rev. Mod. Phys.} {\bf 73}, 977
        (2001).
\bibitem[Szabo(2003)]{noncom} R.J. Szabo, ``Quantum field theory on noncommutative spaces", {\it Phys. Rep.}
        {\bf 378}, 207 (2003).
%
%
\bibitem[Minwalla(2000)]{IRUV} S. Minwalla, M.V. Raamsdonk and N. Seiberg, ``Noncommutative perturbative dynamics", {\it J. High Energy
        Phys.} {\bf 02}, 020 (2000).
\bibitem[Gross(2000)]{TRN} D.J. Gross and N.A. Nekrasov, ``Dynamics of strings in noncommutative gauge theory", {\it J. High
        Energy Phys.} {\bf 10}, 021 (2000).
\bibitem[Szabo(2001)]{TRN1}  F. Lizzi, R.J. Szabo and A. Zampini, ``Geometry of the gauge algebra in noncommutative Yang--Mills
        theory", {\it J. High Energy Phys.} {\bf 08} ,032 (2001).
\bibitem{HO1}  J. Gamboa, M. Loewe and J.C. Rojas, ``Non--commutative quantum mechanics", {\it Phys. Rev. D}
        {\bf 64}, 067901 (2001).

\bibitem{LV} S.M. Carroll, J.A. Harvey, V.A. Kostelecky, C.D. Lane and T. Okamoto, ``Noncommutative field theory and
        Lorentz violation", {\it Phys. Rev. Lett.} {\bf 87}, 141601 (2001).
\bibitem{LV2} A. Anisimov, T. Banks, M. Dine and M. Graesser, ``Comments on non-commutative phenomenology",
        {\it Phys. Rev. D} {\bf 65}, 085032 (2002).
\bibitem{HO4} B. Muthukumar and P. Mitra, ``Non-commutative oscillators and the commutative limit", {\it Phys. Rev.
        D} {\bf 66}, 027701 (2002).
\bibitem{LV3} J.M. Carmona, J.L. Cort\'{e}s, J. Gamboa and F. M\'{e}ndez, ``Noncommutativity in field space and
        Lorentz invariance violation", {\it Phys. Lett. B} {\bf 565}, 222 (2003).
%
%
\bibitem{phrev3} G. Amelino-Camelia, G. Mandanici, and K. Yoshida, ``On the IR/UV mixing and experimental limits
        on the parameters of canonical noncommutative spacetimes", {\it J. High Energy Phys.} {\bf 0401},
        037 (2004).
\bibitem{HAS1} X. Calmet, ``What are the bounds on space-time noncommutativity?", {\it Eur. Phys. J. C}
        {\bf 41}, 269 (2005).
\bibitem{GW} O. Bertolami, J.G. Rosa, C.M.L. de Aragao, P. Castorina and D. Zappala, ``Noncommutative gravitational
        quantum well", {\it Phys. Rev. D} {\bf 72}, 025010 (2005).
\bibitem{GW1} R. Banerjee, B. Dutta Roy and S. Samanta, ``Remarks on the noncommutative gravitational quantum well",
        {\it Phys. Rev. D} {\bf 74}, 045015 (2006).

\bibitem{MF12}B. Malekolkalami and M. Farhoudi, ``Noncommutative double scalar fields in FRW
        cosmology as cosmical oscillators", \textit{Class. Quant. Grav.}\ {\bf 27}, 245009 (2010).
\bibitem{RFK11}S. M. M. Rasouli, M. Farhoudi and N. Khosravi, ``Horizon problem remediation via deformed phase space",
        {\it Gen. Rel. Grav.} {\bf 43}, 2895 (2011).
\bibitem{phrev1} G. Amelino-Camelia, ``Quantum spacetime phenomenology",
        {\it Living Rev. Rel.} {\bf 16}, 5 (2013).

\bibitem{RZMM14} S. M. M. Rasouli, A.H. Ziaie, J. Marto and P.M. Moniz,
        ``Gravitational collapse of a homogeneous scalar field in deformed phase space",
        {\it Phys. Rev. D} {\bf 89}, 044028 (2014).
\bibitem{RM14} S. M. M. Rasouli and P.V. Moniz, ``Noncommutative minisuperspace, gravity-driven acceleration and
        kinetic inflation", {\it Phys. Rev. D} {\bf 90}, 083533 (2014).
\bibitem{MF21}B. Malekolkalami and M. Farhoudi, ``Gravitomagnetism and non-commutative geometry", {\it Int. J. Theor. Phys.}
        {\bf 53}, 815 (2014).
\bibitem{RZJM16} S. M. M. Rasouli, A.H. Ziaie, S. Jalalzadeh and P.V. Moniz, ``Non-sinular Brans-Dicke collapse in deformed phase space",
         {\it Annals of Physics} {\bf 375} 154 (2016).
\bibitem{sf2016} N. Saba and M. Farhoudi, ``Noncommutative universe and chameleon field
        dynamics", submitted to journal.

\bibitem{STMT} T. Banks, W. Fischler, S.H. Shenker and L. Susskind, ``M theory as a matrix model: A Conjecture",
        {\it Phys. Rev. D} {\bf 55}, 5112 (1997).
\bibitem{STMT2} A. Connes, M.R. Douglas and A. Schwarz, ``Noncommutative geometry and matrix theory: Compactification
        on Tori", {\it J. High Energy Phys.} {\bf 02}, 003 (1998).
\bibitem{STMT1} N. Seiberg and E. Witten, ``String theory and noncommutative geometry", {\it J. High Energy Phys.}
        {\bf 09}, 032 (1999).
\bibitem{DFR} S. Doplicher, K. Fredenhagen and J.E. Roberts, ``Spacetime quantization induced by classical gravity",
        {\it Phys. Lett. B} {\bf 331}, 39 (1994).
\bibitem{DFR0} S. Doplicher, K. Fredenhagen and J.E. Roberts, ``The quantum structure of spacetime at the Planck
        scale and quantum fields", {\it Commun. Math. Phys.} {\bf 172}, 187 (1995).
\bibitem{DFR1} E.M.C. Abreu, A.C.R. Mendes, W. Oliveira and A.O. Zangirolami, ``The noncommutative
        Doplicher-Fredenhagen-Roberts-Amorim space", {\it SIGMA} {\bf 6}, 083
        (2010).
\bibitem{DFR2} E.M.C. Abreu and M.J. Neves, ``Causality in noncommutative spacetime", arXiv:1108.5133 {[}hep-th{]};
        \textit{ibid} arXiv:1310.8352.



        \bibitem{inf00}R. Brandenberger and P.M. Ho, ``Noncommutative spacetime, stringy spacetime uncertainty principle, and density fluctuations", {\it Phys. Rev. D} {\bf 66}, 023517 (2002).
\bibitem{gu}A.H. Guth, `` The inflationary universe: A possible solution to the horizon and flatness
        problems",  {\it Phys. Rev. D} {\bf 23}, 347 (1981).
\bibitem{li} A.D. Linde, ``A new inflationary universe scenario: A possible solution of the horizon, flatness,
        homogeneity, isotropy and primordial monopole problems", {\it Phys. Lett. B} {\bf 108}, 389 (1982).
\bibitem{al} A. Albrecht and P.J. Steinhardt, ``Cosmology for grand unified theories with radiatively
        induced symmetry breaking", {\it Phys. Rev. Lett.} {\bf 48}, 1220 (1982).
\bibitem{li1} A.D. Linde, ``Chaotic Inflation", {\it Phys. Lett. B} {\bf 129}, 177 (1983).
\bibitem{linde} A.D. Linde, {\it Particle Physics and Inflationary Cosmology}, (Harwood Chur,
        Switzerland, 1990).
\bibitem{wein2}S. Weinberg, {\it Cosmology,} (Oxford University Press, Oxford, 2008).
\bibitem{6} V. Mukhanov and G. Chibisov, ``Quantum fluctuation and nonsingular universe", {\it J. Exper.
        Theor. Phys. Lett.} {\bf 33}, 532 (1981); {\it Sov. Phys. JETP} {\bf 56}, 258 (1982).
     \bibitem{inf4}Q.C. Huang and M. Li, ``Power spectra in spacetime noncommutative Inflation", {\it Nucl.Phys.B}{\bf 713}, 219 (2005).
      \bibitem{inf0}F. Lizzi, G. Mangano, G. Miele and M. Peloso, ``Cosmological perturbations and short distance physics from noncommutative geometry", {\it JHEP} {\bf 049}, 0206 (2002).

\bibitem{inf1}Q.C. Huang and M. Li, ``CMB power spectrum from noncommutative spacetime", {\it JHEP} {\bf 0306}, 014 (2003).






\bibitem{inf2}S. Tsujikawa, R. Maartens and R. Brandenberger, ``Noncommutative inflation and the CMB", {\it Phys. Lett. B} {\bf 574 }, 141 (2003).
\bibitem{inf3}Q.C. Huang and M. Li, ``Noncommutative inflation and the CMB multipoles", {\it JCAP} {\bf 0311}, 001 (2003).
\bibitem{inf5}H. Kim, G.S. Lee and Y.S. Myung, ``Noncommutative spacetime effect on the slow--roll period of
inflation", {\it Mod. Phys. Lett. A} {\bf 20}, 271 (2005).
\bibitem{AMMOO05} E.M.C. Abreu, M.V.Marcial, A.C.R. Mendes, W. Oliveira and G., Oliveira-Neto, {\it JHEP} {\bf 05}, 144
        (2012).
        \bibitem{S80} A.A. Starobinsky,  {\it Phys. Lett. B} {\bf 91} 99 (1980).
 \bibitem{V85}A. Vilenkin,  {\it Phys. Rev. D} {\bf 32} 2511 (1985).
\bibitem{nonocomshrefs1} G. Esposito and C. Stornaiolo, {\it Int. J. Geom. Meth. Mod. Phys.} {\bf 4}, 349

\bibitem{GUPrefs}
        M.V. Battisti and G. Montani, {\it Phys. Lett. B} {\bf 656}, 96 (2007);
        M.V.  Battisti and G. Montani, {\it Phys. Rev. D} {\bf 77},
        023518 (2008).
\bibitem{reviewnoncom1} M.R. Douglas and N.A. Nekrasov, {\it Rev. Mod. Phys.} {\bf 73}, 977 (2001);\\
        R.J. Szabo, {\it Phys. Rep.} \textbf{378}, 207 (2003).
\bibitem{minwramsei} S. Minwalla, M. Van Raamsdonk and N. Seiberg, {\it JHEP} {\bf 02}, 020 (2000);
        M. Van Raamsdonk and N. Seiberg, {\it JHEP} {\bf 03}, 035 (2000); S. Minwalla, M. Van Raamsdonk and N. Seiberg, {\it
        JHEP} {\bf 0002}, 020 (2000); A. Micu and M.M. Sheikh-Jabbari, {\it JHEP} {\bf
        0101}, 025 (2001).

 \bibitem{GSS11} W. Guzman, M. Sabido and J. Socorro, {\it Phys. Lett. B} {\bf 697}, 271 (2011).

 \bibitem{SS17} C. P. Singh and M. Srivastava, {\it Pramana J. Phys.} {\bf 88}, 22 (2017).

\bibitem{Planck-2015-1}Planck collaboration, P.A.R. Ade et al., {\it Planck 2015 results. XIII. Cosmological
parameters}, Astron. Astrophys. {\bf 594}, A13 (2016)  [arXiv:1502.01589].

\bibitem{Planck-2015-2} Planck collaboration, P.A.R. Ade et al., {\it Planck 2015 results. XX. Constraints on inflation},
Astron. Astrophys. {\bf 594}, A20 (2016) [arXiv:1502.02114].




\bibitem{SL93} E.D. Stewart and D.H. Lyth, ``A more accurate analytic calculation of the spectrum of cosmological
        perturbations produced during inflation", {\it Phys. Lett. B} {\bf 302}, 171 (1993).
\bibitem{MS03} J. Martin and D.J. Schwarz, ``WKB approximation for inflationary cosmological perturbations",
        {\it Phys. Rev. D} {\bf 67}, 083512 (2003).
\bibitem{GS01} J.O. Gong and E.D. Stewart, ``The density perturbation power spectrum to second-order corrections
        in the slow-roll expansion", {\it Phys. Lett. B} {\bf 510}, 1 (2001).
\bibitem{CF05} R. Casadio, F. Finelli, M. Luzzi and G. Venturi, ``Improved WKB analysis of slow-roll inflation",
        {\it Phys. Rev. D} {\bf 72}, 103516 (2005).
\bibitem{Tasi} D. Baumann, ``TASI lectures on inflation'', {\it arXiv: 0907.5424}.
\bibitem{weltman1} J. Khoury and A. Weltman, ``Chameleon cosmology", {\it Phys. Rev. D} {\bf 69},
        044026 (2004).
\bibitem{gubser} S.S. Gubser and J. Khoury, ``Scalar self-interactions loosen constraints from
        fifth force searches", {\it Phys. Rev. D} {\bf 70}, 104001 (2004).
%

\bibitem{saba} N. Saba and M. Farhoudi, ``Chameleon field dynamics during inflation", to appear in {\it Int. J. Mod. Phys. D}, arXiv:1711.09682.

\bibitem{B90} J.D. Barrow, ``The behavior of intermediate inflationary universes", {\it Phys. Lett. B}
        {\bf 249}, 406 (1990).
\bibitem{M02} S. Mukohyama, ``Brane cosmology driven by the rolling tachyon", {\it Phys. Rev. D} {\bf 66}, 024009 (2002).
\bibitem{GL96} A.M. Green and A.R. Liddle, ``Conditions for successful extended inflation", {\it Phys. Rev. D}
        {\bf 54}, 2557 (1996).

\bibitem{LM85} F. Lucchin and S. Matarrese, ``Power-law inflation", {\it Phys. Rev. D}
        {\bf 32}, 1316 (1985).

\bibitem{JH87} J.J. Halliwell, ``Scalar fields in cosmology with an exponential potential", {\it Phys. Lett. B}
        {\bf 185}, 341 (1987).

\bibitem{YM88} J. Yokoyama, K. Maeda, ``On the dynamics of the power law inflation due to an exponential potential", {\it Phys. Lett. B}
        {\bf 207}, 31 (1988).

\bibitem{US13}S. Unnikrishnan and V. Sahni, {\it JCAP}  {\bf 10}, 063 (2016).


%
%
\end{thebibliography}
\end{document}